\documentclass[aps,pra,preprint,superscriptaddress,nofootinbib,longbibliography]{revtex4-2}

\usepackage{amsmath}
\usepackage{amssymb}
\usepackage{bm}
\usepackage{graphicx}
\makeatletter
\def\frontmatter@abstract@produce{%
  \par
  \addvspace{\frontmatter@preabstractspace}%
   \begingroup
    \dimen@\baselineskip
    \setbox\z@\vtop{\unvcopy\absbox}%
    \advance\dimen@-\ht\z@\advance\dimen@-\prevdepth
    \@ifdim{\dimen@>\z@}{\vskip\dimen@}{}%
   \endgroup
   \begingroup
    \prep@absbox
    \unvbox\absbox
    \post@absbox
   \endgroup
  \@ifx{\@empty\mini@notes}{}{\mini@notes\par}%
  \addvspace\frontmatter@postabstractspace
}%
\makeatother

\begin{document}

\title{Quantum and classical thermal equilibrium for\\ probabilistic classical fields}

\author{C.~Wetterich}
\affiliation{Institut f\"ur Theoretische Physik, Universit\"at Heidelberg, Philosophenweg 16, 69120 Heidelberg, Germany\\ c.wetterich@thphys.uni-heidelberg.de}

\begin{abstract}
The time evolution of the probability distribution for classical fields can be described by the quantum formalism. For free classical fields the free quantum field theory arises as a subsystem. Different probability distributions for classical fields correspond to the classical thermal equilibrium with the Rayleigh-Jeans divergence or the quantum thermal equilibrium with the Planck spectrum. The selection of the Rayleigh-Jeans or Planck spectrum is not a question of fundamental concepts. It becomes an issue of dynamical selection once interactions are present. We compare the momentum dependence of the classical energy and field correlations for the classical and quantum equilibrium. This permits to decide which equilibrium is selected as a result of asymptotic dynamics by numerical simulations of the probabilistic classical fields including interactions. For free probabilistic classical fields we construct a general ``classical-quantum map'' between the classical probability distribution and the density matrix of the quantum field theory.
\end{abstract}

\maketitle

\section{Introduction}

Quantum physics has started with an apparent failure of classical physics. The classical thermal equilibrium distribution for electromagnetic fields according to the Boltzmann law shows an ultraviolet divergence -- the Rayleigh-Jeans divergence -- in contrast to observation. Planck's ad hoc suggestion of discrete photons has cured this divergence, predicting the well behaved Planck spectrum for black body electromagnetic radiation. Its temperature behavior has been confirmed with high precision in the cosmic microwave background radiation \cite{CMB1,CMB2}. Within the later developed formulation of quantum physics and quantum field theory the Planck spectrum follows as a natural consequence \cite{TFT1,TFT2,TFT3}. This seems to be a strong argument that quantum physics is fundamentally different from classical physics.

In this note we argue that this last conclusion is premature. For the example of a free complex classical scalar field we describe a classical probability distribution for field configurations which produces the Planck spectrum. This distribution is homogeneous and static, and this ``quantum thermal equilibrium'' depends on a temperature parameter $T$. It is a valid candidate for a state to which a system may relax for large time once suitable small interactions are added. The distinction between Rayleigh-Jeans and Planck needs no fundamental conceptual difference. It becomes a distinction between two classical probability distributions which could describe thermal equilibrium. Our basic concept is a probability distribution for classical fields, not the classical approximation to a quantum field theory \cite{PSTA,DTS,YK,GLV,FGL}. This matters once interactions are introduced at a later stage such that the dynamics may decide which equilibrium is reached as an asymptotic state for large time. In contrast to stochastic electrodynamics \cite{MAR1,BOY1,BOY2,BOY3,QDICE}, we do not couple the classical field to stochastic sources.

In the zero temperature limit the classical probability distribution for the classical thermal equilibrium becomes a sharp $\delta$-distribution, which singles out a sharp zero field value for the ground state. In contrast, the classical probability distribution for the quantum thermal equilibrium state becomes in the zero temperature limit a Gaussian with a finite width. In the quantum ground state the field is fluctuating. This is in accordance with the proposal to use fluctuating field observables for a more robust description of statistical properties of classical fields \cite{CWQF}. For non-zero temperature the classical thermal equilibrium is described by a Gaussian with width given by $T$.

The classical probability distribution for the quantum thermal equilibrium corresponds precisely to the thermal equilibrium distribution for a quantum field theory for a free complex scalar field. It is closely related to the possibility of obtaining a quantum field theory as a subsystem of a classical probability field theory \cite{CWQF}. It is well known that free quantum field theories and free classical field theories share many common features. The generating functionals for equal-time correlation functions (equal-time quantum effective action) obey the same evolution equation \cite{CWET1,CWET2}. In the Schwinger-Keldysh formulation for the quantum field theory the classical approximation becomes exact \cite{BER,KAM} for a free field theory, which permits numerical simulations of the quantum field theory \cite{AS2,CKH,EGW,AS1,BUJA,BBG} in terms of classical fields. For the thermal equilibrium of a free quantum field theory the Wigner function is positive and can define a probability distribution.

The fact that a classical probability distribution can describe the Planck spectrum is known since a long time. What is new in the present paper is the particular setting how a probability distribution for classical fields can describe a quantum field theory. The thermal equilibrium state of a free quantum field theory is a particular state for which many different formulations of a classical probability distribution can be employed. This changes if one considers non-equilibrium quantum states and includes interactions. The quantum formalism for probabilistic classical fields employed here extends to arbitrary solutions of the Liouville equation for probabilistic classical fields, including interactions. The free field theory still has special properties related to the existence of a particularly simple closed quantum subsystem -- namely the free quantum field theory. Nevertheless, the generic quantum features extend beyond the free fields. This concerns, in particular, the presence of statistical observables which are represented by non-commuting operators in the quantum formalism -- a feature that is typically not present in other approaches.

In the present paper we focus on the thermal equilibrium states in this approach to probabilistic classical fields. The same setting can describe the classical thermal equilibrium with the Rayleigh-Jeans divergence and the quantum thermal equilibrium which yields the Planck spectrum. They can therefore be compared, which is a crucial ingredient if one later wants to judge if in the presence of interactions the dynamics prefers the classical or quantum equilibrium as an asymptotic state. We note that in our formalism the classical probability distribution for the quantum thermal equilibrium is not the Wigner function, which is usually used in numerical simulations. This comes in pair with the fact that the energy distribution of the Planck spectrum concerns the quantum energy, which differs from the classical energy. It is the quantum energy for which the harmonic oscillator shows a discrete energy spectrum, and the corresponding quantum Hamiltonian leads to the Schr\"odinger equation for a particle in an arbitrary potential if inhomogeneities are included for the free theory \cite{CWDS}. For a free theory we construct a ``quantum-classical map'' between a classical probability distribution and a quantum density matrix for a quantum subsystem. This permits us to compute both the quantum energy and the classical energy for both the quantum and the classical thermal equilibrium state and to proceed to detailed comparisons.

A basic ingredient is the real classical wave function \cite{CWQP1,CWIT,CWQFC} which is the root of the probability distribution. It differs from the complex Koopman--von Neumann--Sch\"onberg wave function \cite{KOOP,VNEU,SCHB}. The latter is used for many interesting developments in classical statistical systems \cite{KAN,MAMA,MAU,GOMA,NIC,NAKL,VOLO,BON,AME,KPM,ACKA,CHRU,KLE,MEZI,DW,GOZ1,GRT}. It cannot describe quantum mechanics, however, since phases are redundant. A second key concept are statistical observables \cite{CWQF,CWQP,CWDS}. They are represented by non-commuting operators, and their correlations do not need to obey Bell's inequalities \cite{BEL1,CHSH}. The starting point of our investigation is a positive probability distribution for classical fields. It differs from well known positive probability distributions in quantum optics based on the Husimi function \cite{HUS} or the Glauber-Sudarshan $P$-function \cite{GLA,SUD,GLA2,CAGL}, which are based on the coherent states of a quantum field theory.

We start in sect.~\ref{sec:harmonic} by a discussion of the probabilistic classical harmonic oscillator. We describe the quantum subsystem which yields the quantum harmonic oscillator with its discrete spectrum for the quantum energy. For this simple system we compare the quantum and classical energy. In sect.~\ref{sec:fieldtheory} we turn to the probabilistic classical field theory. It can be seen as an ensemble of classical harmonic oscillators for the different momentum modes. At this point one understands why the Rayleigh-Jeans divergence arises for the classical thermal equilibrium, but not for the quantum thermal equilibrium. In sect.~\ref{sec:qc-connection} we discuss in more detail the connection between a classical probabilistic field theory and its subsystem described by a quantum field theory. We discuss a general classical-quantum map and compare our approach with the positive probability distributions used in quantum optics. We extend the discussion to correlation functions evolving in time and in the presence of interactions. We discuss our results in sect.~V.

\section{Probabilistic classical harmonic oscillator}
\label{sec:harmonic}

The difference between the quantum and classical thermal equilibrium states can already be seen on the level of a harmonic oscillator. Since this will be the basis for a free field theory we begin our investigation with this simple case.

\subsection{Harmonic oscillator}
\label{sec:ho}

The classical evolution equation for a (one-dimensional) harmonic oscillator reads
\begin{equation}
\partial_t z = v \,, \qquad \partial_t v = -\omega^2 z \,.
\label{eq:zv-evolution}
\end{equation}
with the conserved classical energy
\begin{equation}
E^{(cl)}(z,v) = \frac{1}{2} v^2 + \frac{\omega^2}{2} z^2 \,.
\label{eq:E-cl-ho-def}
\end{equation}
The corresponding Liouville equation for the probability distribution $w(z,v)$ in phase space is given by
\begin{equation}
\begin{aligned}
\partial_t w &= -\hat L\, w \,, \\[2pt]
\hat L &= v\,\frac{\partial}{\partial z} - \omega^2 z\, \frac{\partial}{\partial v} \,.
\end{aligned}
\label{eq:Liouville}
\end{equation}
We employ the real classical wave function $q(z,v)$, which is the root of $w(z,v)$ \cite{CWQP1,CWIT,CWQFC},
\begin{equation*}
w(z,v) = q^2(z,v) \,.
\end{equation*}
Its evolution follows the Liouville equation as well. It can be written in Hamiltonian form
\begin{equation}
i\,\partial_t q = H_L\, q \,, \qquad H_L = -i \hat L \,.
\label{eq:HL-def}
\end{equation}
The hermitian Hamiltonian operator $H_L$ can be expressed in terms of the hermitian operators
\begin{equation}
\begin{aligned}
\hat z &= z \,, & \hat \gamma &= -i \frac{\partial}{\partial z} \,, \\
\hat v &= v \,, & \hat\zeta &= i \frac{\partial}{\partial v} \,,
\end{aligned}
\label{eq:hat-ops}
\end{equation}
as
\begin{equation}
H_L = \hat v \, \hat \gamma + \omega^2 \hat z \, \hat\zeta \,.
\label{eq:HL-explicit}
\end{equation}

One can decompose $H_L$ into two independent pieces by suitable linear combinations of operators \cite{CWQP}
\begin{equation}
\begin{aligned}
\hat x &= \hat z + \tfrac12 \hat\zeta \,, & \hat y &= \hat z - \tfrac12 \hat\zeta \,, \\
\hat \eta &= \hat v + \tfrac12 \hat\gamma \,, & \hat \beta &= \hat v - \tfrac12 \hat\gamma \,,
\end{aligned}
\label{eq:new-ops}
\end{equation}
namely
\begin{equation}
\begin{aligned}
H_L &= H^{(q)} - H^{(m)} \,, \\
H^{(q)} &= \tfrac12 \hat\eta^2 + \tfrac{\omega^2}{2} \hat x^2 \,, \\
H^{(m)} &= \tfrac12 \hat\beta^2 + \tfrac{\omega^2}{2} \hat y^2 \,.
\end{aligned}
\label{eq:HL-decomp}
\end{equation}
One observes the commutation relations
\begin{equation}
[\hat x, \hat \eta] = i \,, \qquad [\hat y, \hat \beta] = -i \,,
\label{eq:commutators}
\end{equation}
while the operators $(\hat x,\hat\eta)$ commute with the operators $(\hat y,\hat\beta)$. The operators $(\hat x,\hat\eta)$ describe a quantum harmonic oscillator. The ``mirror particle'' described by $(\hat y,\hat\beta)$ is related to it by time reversal.

An important advantage of the use of wave functions is the possibility to perform basis transformations. We define the complex wave function
\begin{equation}
\tilde\psi(z,\zeta) = \int_v \exp(i v \zeta)\, q(z,v) \,,
\label{eq:psi-tilde-def}
\end{equation}
with $\int_v = (2\pi)^{-1}\int_{-\infty}^{\infty} dv$, and correspondingly $\tilde\psi(x,y)$ with $x = z + \zeta/2$, $y = z - \zeta/2$. In this basis the operators are represented as
\begin{equation}
\begin{aligned}
\hat x &= x \,, & \hat \eta &= -i \frac{\partial}{\partial x} \,, \\
\hat y &= y \,, & \hat \beta &= i \frac{\partial}{\partial y} \,,
\end{aligned}
\label{eq:ops-xy-basis}
\end{equation}
and
\begin{equation}
H^{(q)} = -\tfrac12 \partial_x^2 + \tfrac{\omega^2}{2} x^2 \,, \qquad
H^{(m)} = -\tfrac12 \partial_y^2 + \tfrac{\omega^2}{2} y^2 \,.
\label{eq:H-xy}
\end{equation}
Since $q$ is real, the complex wave function $\tilde\psi$ obeys the constraint
\begin{equation}
\tilde\psi^*(z,\zeta) = \tilde\psi(z,-\zeta) \,, \qquad
\tilde\psi^*(x,y) = \tilde\psi(y,x) \,.
\label{eq:reality-constraint}
\end{equation}
Both $H^{(q)}$ and $H^{(m)}$ correspond to the quantum Hamiltonian for the harmonic oscillator.

\subsection{Quantum density matrix}
\label{sec:densitymatrix}

Let us expand an arbitrary complex function of $x$ in terms of some complete orthonormal set of basis functions, $\psi(x) = c_m \psi_m(x)$ (with summation over double indices implied). We can then write the wave function $\tilde\psi(t,x,y)$ as a double expansion
\begin{equation}
\tilde\psi(t,x,y) = b_{mn}(t)\, \psi_m(x)\, \psi_n^*(y) \,.
\label{eq:psi-expansion}
\end{equation}
The condition $\tilde\psi^*(t,x,y) = \tilde\psi(t,y,x)$ implies that the coefficients $b_{mn}(t)$ are elements of a hermitian matrix $b$,
\begin{equation}
b^\dagger = b \,, \qquad b_{mn}^* = b_{nm} \,,
\label{eq:b-hermitian}
\end{equation}
and the normalization of $\tilde\psi$ enforces
\begin{equation}
\operatorname{tr} b^2 = 1 \,.
\label{eq:trace-normalization}
\end{equation}
The matrix $b^2$ is a positive hermitian matrix with all eigenvalues larger or equal to zero. It has the properties of a density matrix in quantum mechanics. In position space we can write the (quantum) density matrix as
\begin{equation}
\rho(t,x,y) = b^2_{mn}(t)\, \psi_m(x)\, \psi_n^*(y) \,,
\label{eq:rho-def}
\end{equation}
with $b^2_{mn}$ the elements of the matrix $b^2$,
\begin{equation}
b^2_{mn} = \sum_q b_{mq}\, b_{qn} \,.
\label{eq:b2-def}
\end{equation}
The density matrix $\rho$ is the square of the wave function $\tilde\psi$ in a matrix sense,
\begin{equation}
\rho = \tilde\psi^2 \,, \qquad \rho(x,x') = \int_y \tilde\psi(x,y)\, \tilde\psi(y,x') \,.
\label{eq:rho-psi-square}
\end{equation}
One can diagonalize $b$ by a suitable similarity transformation with real diagonal elements $\alpha_m$,
\begin{equation}
\tilde b_{mn} = \alpha_m\, \delta_{mn} \,, \qquad
\tilde b^2_{mn} = w_m\, \delta_{mn} \,, \qquad
w_m = \alpha_m^2 \geq 0 \,.
\label{eq:b-diagonal}
\end{equation}
With $\psi'_m(x)$ the similarity transform of $\psi_m(x)$ one has
\begin{equation}
\rho(t,x,y) = \sum_m w_m\, \psi'_m(x)\, \psi_m'^*(y) \,, \qquad \sum_m w_m = 1 \,.
\label{eq:rho-diagonal}
\end{equation}
This is the standard expression of a quantum density matrix as a sum of pure state density matrices with weights $w_m$. Thus $\rho(t,x,y)$ has all the properties of a quantum density matrix in position space. Eq.~\eqref{eq:rho-def} is the expansion of the most general density matrix. In the diagonal form the wave function $\tilde\psi(t,x,y)$ is easily found,
\begin{equation}
\tilde\psi(t,x,y) = \sum_m \alpha_m\, \psi'_m(x)\, \psi_m'^*(y) \,.
\label{eq:psi-diagonal}
\end{equation}

A particular case are pure quantum states for which only one $\alpha_m$ differs from zero,
\begin{equation}
\tilde\psi(t,x,y) = \psi(t,x)\, \psi^*(t,y) \,.
\label{eq:pure-state}
\end{equation}
The evolution of the quantum wave function $\psi(t,x)$ follows the Schr\"odinger equation with Hamiltonian $H^{(q)}$,
\begin{equation}
i\, \partial_t \psi = H^{(q)} \psi \,.
\label{eq:schroedinger}
\end{equation}
The eigenstates of $H^{(q)}$ obey
\begin{equation}
H^{(q)} \psi_m = E_m \psi_m = \left(m+\tfrac12\right)\omega\, \psi_m \,.
\label{eq:Hq-eigen}
\end{equation}
The ground state wave function reads
\begin{equation}
\psi_0 = \left(\frac{\omega}{\pi}\right)^{1/4} \exp\!\left(-\frac{\omega x^2}{2}\right) \,,
\label{eq:psi0}
\end{equation}
which corresponds to
\begin{equation}
\tilde\psi_0 = \left(\frac{\omega}{\pi}\right)^{1/2} \exp\!\left(-\frac{\omega}{2}(x^2+y^2)\right)
= \left(\frac{\omega}{\pi}\right)^{1/2} \exp\!\left(-\omega\left(z^2+\frac{\zeta^2}{4}\right)\right) \,.
\label{eq:psi0tilde}
\end{equation}
This results in the real wave function for the quantum vacuum
\begin{equation}
q_0(z,v) = 2\exp\!\left(-\frac{2}{\omega} E^{(cl)}(z,v)\right) \,,
\label{eq:q0}
\end{equation}
and the quantum vacuum probability distribution
\begin{equation}
w_0(z,v) = 4\exp\!\left(-\frac{4}{\omega} E^{(cl)}(z,v)\right)
= 4\exp\!\left(-\frac{2v^2}{\omega}\right)\exp\!\left(-2\omega z^2\right) \,.
\label{eq:w0}
\end{equation}
This classical probability distribution is a Gaussian, with widths for $z$ and $v$ related and determined by $\omega$.

\subsection{Quantum thermal equilibrium}
\label{sec:qte}

Consider first the quantum thermal equilibrium at temperature $T$. In this case the diagonal density matrix is given by
\begin{equation}
\rho(x,y) = \frac{1}{Z}\sum_m \exp(-\beta E_m)\, \psi_m(x)\, \psi_m^*(y) \,, \qquad
Z = \sum_m \exp(-\beta E_m) \,,
\label{eq:rho-thermal}
\end{equation}
with
\begin{equation}
E_m = \left(m+\tfrac12\right)\omega \,, \qquad \beta = T^{-1} \,,
\label{eq:Em-def}
\end{equation}
and $\psi_m(x)$ the normalized eigenfunctions to the quantum energy $E_m$. The Schr\"odinger equation \eqref{eq:schroedinger}, \eqref{eq:Hq-eigen} for $\psi_m$ implies that this density matrix is static. We can compute the associated classical probability distribution $w(z,v)$. With $b_{mn} = Z^{-1/2}\exp(-\beta E_m/2)\,\delta_{mn}$ one has
\begin{equation}
\tilde\psi(x,y) = \sum_m Z^{-1/2} \exp\!\left(-\tfrac{\beta}{2} E_m\right) \psi_m(x)\, \psi_m^*(y) \,,
\label{eq:psi-thermal-xy}
\end{equation}
or
\begin{equation}
\tilde\psi(z,\zeta) = \sum_m Z^{-1/2} \exp\!\left(-\tfrac{\beta}{2} E_m\right) \psi_m\!\left(z+\tfrac{\zeta}{2}\right) \psi_m^*\!\left(z-\tfrac{\zeta}{2}\right) \,.
\label{eq:psi-thermal-zs}
\end{equation}
Performing the inverse Fourier transformation one obtains
\begin{equation}
q(z,v) = \sum_m Z^{-1/2} \exp\!\left(-\tfrac{\beta}{2} E_m\right) \int_\zeta \exp(-i\zeta v)\, \psi_m\!\left(z+\tfrac{\zeta}{2}\right) \psi_m^*\!\left(z-\tfrac{\zeta}{2}\right) \,.
\label{eq:q-thermal}
\end{equation}
The classical probability distribution $w(z,v)$ which describes the quantum thermal equilibrium state is then found as $w(z,v) = q^2(z,v)$.
For $\beta\to\infty$ only the term $m=0$ contributes, and the wave function and probability distribution approach the Gaussian \eqref{eq:q0}, \eqref{eq:w0} for the quantum ground state. For finite $\beta$ the contributions from the individual modes $m>0$ are no longer Gaussian. Nevertheless, an explicit evaluation of the sum \eqref{eq:psi-thermal-xy} yields a Gaussian
\begin{equation}
\tilde\psi(x,y) = \sqrt{\frac{\omega}{\pi}}\, \exp\!\left\{ -\frac{\omega}{2}\left[ \coth\!\left(\frac{\beta\omega}{2}\right)(x^2+y^2) - \frac{2xy}{\sinh\!\left(\beta\omega/2\right)} \right] \right\} \,.
\label{eq:psi-gauss}
\end{equation}
This results in the Gaussian distributions
\begin{equation}
q(z,v) = 2\sqrt{\tanh\!\left(\frac{\beta\omega}{4}\right)}\, \exp\!\left\{ -\tanh\!\left(\frac{\beta\omega}{4}\right)\left(\omega z^2 + \frac{v^2}{\omega}\right) \right\} \,,
\label{eq:q-gauss}
\end{equation}
and
\begin{equation}
w(z,v) = 4\tanh\!\left(\frac{\beta\omega}{4}\right) \exp\!\left\{ -\frac{4}{\omega}\tanh\!\left(\frac{\beta\omega}{4}\right) E^{(cl)}(z,v) \right\} \,.
\label{eq:w-boltz}
\end{equation}
The latter is a Boltzmann distribution with ``effective temperature''
\begin{equation}
T_{\rm eff} = \frac{\omega}{4} \coth\!\left(\frac{\beta\omega}{4}\right) \,.
\label{eq:Teff}
\end{equation}

We may evaluate the expectation value for both the classical and the quantum energy for this quantum thermal state. Both for classical and statistical observables the expectation values are expressed as bilinear in the classical wave function $q$ \cite{CWDS}. The quantum energy $E^{(q)}$ is represented by the operator $H^{(q)}$, and its expectation value obeys
\begin{equation}
\begin{aligned}
\langle E^{(q)} \rangle_q &= q^T H^{(q)} q = \tilde\psi^\dagger H^{(q)} \tilde\psi \\
&= \int_{x,y} \tilde\psi^*(x,y) \left(-\tfrac12 \partial_x^2 + \tfrac{\omega^2}{2} x^2\right) \tilde\psi(x,y) \\
&= Z^{-1} \sum_m E_m \exp(-\beta E_m) = -\frac{\partial}{\partial\beta} \ln Z \,.
\end{aligned}
\label{eq:Eq-expect}
\end{equation}
This is the standard relation for quantum mechanics. With
\begin{equation}
Z(\beta) = \exp\!\left(-\frac{\beta\omega}{2}\right) \left(1-\exp(-\beta\omega)\right)^{-1} \,,
\label{eq:Z-beta}
\end{equation}
one has
\begin{equation}
\langle E^{(q)} \rangle_q = \omega \left( \frac12 + \frac{\exp(-\beta\omega)}{1-\exp(-\beta\omega)} \right)
= \frac{\omega}{2} \coth\!\left(\frac{\beta\omega}{2}\right) \,,
\label{eq:Eq-explicit}
\end{equation}
with the limits
\begin{equation}
\langle E^{(q)} \rangle_q =
\begin{cases}
\dfrac{\omega}{2} + \omega \exp(-\beta\omega) & \text{for } \beta\omega \gg 1 \,, \\[2mm]
\beta^{-1} & \text{for } \beta\omega \ll 1 \,.
\end{cases}
\label{eq:Eq-limits}
\end{equation}

The expectation value of the classical energy can be evaluated from the wave function of the quantum equilibrium state by use of the operator
\begin{equation}
\begin{aligned}
\hat E^{(cl)} &= \tfrac12 \hat v^2 + \tfrac{\omega^2}{2} \hat z^2 \\
&= \tfrac14 \left( H^{(q)} + H^{(m)} + \hat\eta\hat\beta + \omega^2 \hat x \hat y \right) \,,
\end{aligned}
\label{eq:Ecl-op}
\end{equation}
One finds
\begin{equation}
\langle E^{(cl)} \rangle_q = q^T \hat E^{(cl)} q = \tfrac12 \langle E^{(q)} \rangle_q + \Delta_E \,,
\label{eq:Ecl-expect}
\end{equation}
with
\begin{equation}
\Delta_E = \tfrac14 \tilde\psi^\dagger \left(\hat\eta\hat\beta + \omega^2 \hat x \hat y\right) \tilde\psi
= \frac{1}{4Z} \sum_{mn} \exp\!\left(-\tfrac{\beta}{2}(E_m+E_n)\right) \left[ \omega^2 |x_{mn}|^2 + |p_{mn}|^2 \right] \,.
\label{eq:DeltaE-def}
\end{equation}
The transition elements,
\begin{equation}
\begin{aligned}
x_{mn} &= \int_x \psi_m^* \, x\, \psi_n = \frac{1}{\sqrt{2\omega}} \langle m | (a^\dagger + a) | n \rangle \,, \\
p_{mn} &= \int_x \psi_m^* \left(-i\frac{\partial}{\partial x}\right) \psi_n = i\sqrt{\frac{\omega}{2}} \langle m | (a^\dagger - a) | n \rangle \,,
\end{aligned}
\label{eq:xmn-pmn}
\end{equation}
are expressed in terms of the creation and annihilation operators obeying
\begin{equation}
a|n\rangle = \sqrt{n}\,|n-1\rangle \,, \qquad a^\dagger|n\rangle = \sqrt{n+1}\,|n+1\rangle \,, \qquad [a,a^\dagger] = 1 \,.
\label{eq:ladder-ops}
\end{equation}
One obtains
\begin{equation}
\omega^2 |x_{mn}|^2 = \frac{\omega}{2}\left[(n+1)\delta_{m,n+1} + n\,\delta_{m,n-1}\right] \,, \qquad
|p_{mn}|^2 = \frac{\omega}{2}\left[(n+1)\delta_{m,n+1} + n\,\delta_{m,n-1}\right] \,,
\label{eq:xmn-pmn-explicit}
\end{equation}
and
\begin{equation}
\begin{aligned}
\Delta_E &= \frac{\omega}{4Z} \sum_n \left\{ (n+1)\exp\!\left(-\tfrac{\beta}{2}(E_n+E_{n+1})\right) + n\exp\!\left(-\tfrac{\beta}{2}(E_n+E_{n-1})\right) \right\} \\
&= \tfrac12 \langle E^{(q)} \rangle_q \cosh\!\left(\frac{\beta\omega}{2}\right) - \frac{\omega}{4} \sinh\!\left(\frac{\beta\omega}{2}\right)
= \frac{\omega}{4\sinh\!\left(\beta\omega/2\right)} \,.
\end{aligned}
\label{eq:DeltaE-explicit}
\end{equation}
and therefore
\begin{equation}
\langle E^{(cl)} \rangle_q = \frac{\omega}{4} \coth\!\left(\frac{\beta\omega}{4}\right) \,.
\label{eq:Ecl-q-explicit}
\end{equation}
One may also evaluate $\langle E^{(cl)} \rangle_q$ directly from the Boltzmann distribution \eqref{eq:w-boltz}. The result $\langle E^{(cl)} \rangle_q = T_{\rm eff}$, with $T_{\rm eff}$ given by eq.~\eqref{eq:Teff}, conforms with our evaluation in the quantum formalism, eq.~\eqref{eq:Ecl-q-explicit}. Comparing eqs.~\eqref{eq:Ecl-q-explicit} and \eqref{eq:Eq-explicit} the expectation value of the classical energy is obtained from the one of the quantum energy by $\omega \to \omega/2$.

In the high temperature limit, $\beta\omega \ll 1$, the expectation values of $E^{(q)}$ and $E^{(cl)}$ approach each other,
\begin{equation}
\langle E^{(cl)} \rangle_q = \langle E^{(q)} \rangle_q - \frac{\beta\omega^2}{16} + \ldots
\label{eq:Ecl-small-betaomega}
\end{equation}
For $\beta\omega \gg 1$ one finds
\begin{equation}
\Delta_E = \frac{\omega}{2} \exp\!\left(-\frac{\beta\omega}{2}\right) \,,
\label{eq:DeltaE-large-betaomega}
\end{equation}
\begin{equation}
\langle E^{(cl)} \rangle_q = \frac{\omega}{4} + \frac{\omega}{2} \exp\!\left(-\frac{\beta\omega}{2}\right) \,.
\label{eq:Ecl-q-large-betaomega}
\end{equation}
The structure is similar to $\langle E^{(q)} \rangle_q$, but the ground state energy is reduced by a factor two.

\subsection{Classical thermal equilibrium}
\label{sec:cte}

The classical thermal equilibrium is most easily expressed in terms of the probability distribution $w(z,v)$,
\begin{equation}
w(z,v) = Z_{cl}^{-1} \exp\!\left(-\beta E^{(cl)}(z,v)\right) \,, \qquad
Z_{cl} = \int_{z,v} \exp\!\left(-\beta E^{(cl)}(z,v)\right) \,.
\label{eq:w-classical}
\end{equation}
This distribution is Gaussian for all values of $\beta$. For $\beta\to\infty$, one obtains the classical ground state,
\begin{equation}
\bar w_0(z,v) = \delta(z)\, \delta(v) \,.
\label{eq:w0bar}
\end{equation}
The expectation value of the classical energy $E^{(cl)}$ in the classical thermal equilibrium state reads
\begin{equation}
\langle E^{(cl)} \rangle_{cl} = \int_{z,v} E^{(cl)}(z,v)\, w(z,v) = -\frac{\partial}{\partial\beta} \ln Z_{cl}(\beta) \,.
\label{eq:Ecl-cl-def}
\end{equation}
The Gaussian integral yields
\begin{equation}
Z_{cl}(\beta) = \int_{z,v} \exp\!\left\{-\frac{\beta}{2}\left(v^2+\omega^2 z^2\right)\right\} = \frac{1}{\beta\omega} \,,
\label{eq:Zcl-explicit}
\end{equation}
and
\begin{equation}
\langle E^{(cl)} \rangle_{cl} = \frac{1}{\beta} = T \,.
\label{eq:Ecl-cl-explicit}
\end{equation}
In the high temperature limit $\beta\omega \ll 1$, this equals the quantum thermal equilibrium, $\langle E^{(cl)} \rangle_q = \langle E^{(q)} \rangle_q = T$. The high frequency limit $\beta\omega \gg 1$ differs qualitatively from the value for the quantum thermal state. This difference will lead for the classical field theory to the Rayleigh--Jeans divergence for the classical equilibrium.

For the computation of the quantum energy in the classical thermal equilibrium we employ the wave function,
\begin{equation}
\langle E^{(q)} \rangle_{cl} = q_{cl}^T H^{(q)} q_{cl} = \tilde\psi_{cl}^\dagger H^{(q)} \tilde\psi_{cl} \,.
\label{eq:Eq-cl-def}
\end{equation}
The wave functions in the different bases are given by
\begin{equation}
q_{cl}(z,v) = Z_{cl}^{-1/2} \exp\!\left(-\frac{\beta}{4}\left(v^2+\omega^2 z^2\right)\right) \,,
\label{eq:qcl-zv}
\end{equation}
and
\begin{equation}
\begin{aligned}
\tilde\psi_{cl} &= \sqrt{\frac{\omega}{\pi}} \exp\!\left\{-\left(\frac{\beta\omega^2 z^2}{4}+\frac{\zeta^2}{\beta}\right)\right\} \\
&= \sqrt{\frac{\omega}{\pi}} \exp\!\left\{-\beta\left[\left(\frac{\omega^2}{16}+\frac{1}{\beta^2}\right)(x^2+y^2) + \left(\frac{\omega^2}{8}-\frac{2}{\beta^2}\right) xy\right]\right\} \,.
\end{aligned}
\label{eq:psicl-zzeta}
\end{equation}
For the operator $H^{(q)}$ we employ
\begin{equation}
\begin{aligned}
H^{(q)} &= \frac12\left(\hat v+\frac12\hat\gamma\right)^2 + \frac{\omega^2}{2}\left(\hat z+\frac12\hat\zeta\right)^2 \\
&= \hat E^{(cl)} + \frac18\left(\hat\gamma^2+\omega^2\hat\zeta^2\right) + \frac12 H_L \,.
\end{aligned}
\label{eq:Hq-cl-decomp}
\end{equation}
The classical thermal equilibrium is static,
\begin{equation}
H_L\, q_{cl} = 0 \,.
\label{eq:HLqcl-zero}
\end{equation}
Then the expectation value of the quantum energy in the classical thermal equilibrium exceeds the classical energy by $\langle\Delta_q\rangle_{cl}$,
\begin{equation}
\langle E^{(q)} \rangle_{cl} = \langle E^{(cl)} \rangle_{cl} + \langle\Delta_q\rangle_{cl} \,, \qquad
\langle\Delta_q\rangle_{cl} = \frac18\left(\langle\hat\gamma^2\rangle_{cl}+\omega^2\langle\hat\zeta^2\rangle_{cl}\right) \,.
\label{eq:Eq-cl-split}
\end{equation}
In the basis \eqref{eq:hat-ops} for $q_{cl}(z,v)$ one obtains
\begin{equation}
\langle\Delta_q\rangle_{cl} = \frac18 \int_{z,v} \left[\left(\frac{\partial q_{cl}}{\partial z}\right)^2 + \omega^2 \left(\frac{\partial q_{cl}}{\partial v}\right)^2\right]
= \frac{\beta^2\omega^2}{16} \langle E^{(cl)} \rangle_{cl} = \frac{\beta\omega^2}{16} \,,
\label{eq:Deltaq-cl-explicit}
\end{equation}
and therefore
\begin{equation}
\langle E^{(q)} \rangle_{cl} = \frac{1}{\beta} + \frac{\beta\omega^2}{16} \,.
\label{eq:Eq-cl-explicit}
\end{equation}
In the large temperature limit $\beta\omega \ll 1$ the second term is small and $\langle E^{(q)} \rangle_{cl}$ approaches $\langle E^{(cl)} \rangle_{cl}$. In contrast, in the large frequency limit $\beta\omega \gg 1$ the contribution $\Delta_q$ dominates and $\langle E^{(q)} \rangle_{cl} \gg \langle E^{(cl)} \rangle_{cl}$.

In summary, in the large temperature limit one finds the universal result $\langle E \rangle = T$. This holds both for $E^{(q)}$ and $E^{(cl)}$, and both for the quantum equilibrium and the classical equilibrium. For low temperatures or large frequencies, $\beta\omega \gg 1$, the various energy expectation values differ. Only the classical energy in the classical equilibrium keeps the value $T$ independently of the frequency. For the quantum equilibrium $\langle E^{(cl)} \rangle_q$ approaches a constant $\omega/4$, with exponentially suppressed corrections. In this limit the probability distribution approaches the quantum vacuum \eqref{eq:w0}, for which $\langle E^{(cl)} \rangle_q$ indeed equals $\omega/4$. The key difference between the classical and quantum equilibrium is the suppression of contributions with large $v^2 \gg T$ for the classical equilibrium, and $v^2 \gg \omega$ for the quantum equilibrium independently of $T$. This holds similarly for $\omega^2 z^2$. For $T \ll \omega$ the probability distribution of the quantum equilibrium extends to much larger values of $v^2$ and $\omega^2 z^2$ as compared to the classical equilibrium, and it becomes almost independent of $T$. The contribution of the excited states of the quantum harmonic oscillator is suppressed by factors $\exp(-\beta\omega)$.

For all states the expectation value of the quantum energy exceeds the one for the classical energy by the value of $\langle\Delta_q\rangle_{cl}$ or $\langle\Delta_q\rangle_q$. The statistical observable $\Delta_q$ is associated to the operator
\begin{equation}
\hat\Delta_q = \frac18\left(\hat\gamma^2+\omega^2\hat\zeta^2\right) \,.
\label{eq:Deltaq-op}
\end{equation}
In turn, $\langle\hat\gamma^2\rangle$ measures the roughness of the wave function in $z$,
\begin{equation}
\langle\hat\gamma^2\rangle = \int_{z,v} \left(\frac{\partial q}{\partial z}\right)^2 \,,
\label{eq:gamma2-def}
\end{equation}
while $\langle\hat\zeta^2\rangle$ yields the roughness in the velocity distribution
\begin{equation}
\langle\hat\zeta^2\rangle = \int_{z,v} \left(\frac{\partial q}{\partial v}\right)^2 \,.
\label{eq:zeta2-def}
\end{equation}
For the low temperature limit of the quantum equilibrium one finds
\begin{equation}
\langle\Delta_q\rangle_q = \frac{\omega}{4} \,,
\label{eq:Deltaq-q-lowT}
\end{equation}
while for the classical equilibrium the distribution is much steeper, resulting in a large enhancement factor
\begin{equation}
\langle\Delta_q\rangle_{cl} = \frac{\beta\omega^2}{16} \,.
\label{eq:Deltaq-cl-lowT}
\end{equation}

\subsection{Two-point functions}
\label{sec:twopoint}

One can express the expectation value of the classical and quantum energy by various two-point functions,
\begin{equation}
\langle E^{(cl)} \rangle = \frac12 \langle v^2 \rangle + \frac{\omega^2}{2} \langle z^2 \rangle \,, \qquad
\langle E^{(q)} \rangle = \langle E^{(cl)} \rangle + \frac18 \langle\gamma^2\rangle + \frac{\omega^2}{8} \langle\zeta^2\rangle \,,
\label{eq:Ecl-Eq-twopoint}
\end{equation}
where the expectation values have to be evaluated in the quantum or classical equilibrium, respectively.

For the classical equilibrium one has
\begin{equation}
\langle z^2 \rangle_{cl} = \frac{1}{\beta\omega^2} \,, \qquad \langle v^2 \rangle_{cl} = \frac{1}{\beta} \,, \qquad
\langle\zeta^2\rangle_{cl} = \frac{\beta^2}{4} \langle v^2 \rangle_{cl} = \frac{\beta}{4} \,, \qquad
\langle\gamma^2\rangle_{cl} = \frac{\beta^2\omega^4}{4} \langle z^2 \rangle_{cl} = \frac{\beta\omega^2}{4} \,.
\label{eq:twopoint-cl}
\end{equation}
In contrast, one finds for the quantum equilibrium
\begin{equation}
\begin{aligned}
\langle z^2 \rangle_q &= \frac{1}{4\omega} \coth\!\left(\frac{\beta\omega}{4}\right) \,, &\qquad
\langle v^2 \rangle_q &= \frac{\omega}{4} \coth\!\left(\frac{\beta\omega}{4}\right) \,, \\
\langle\zeta^2\rangle_q &= \frac{1}{\omega} \tanh\!\left(\frac{\beta\omega}{4}\right) \,, &\qquad
\langle\gamma^2\rangle_q &= \omega \tanh\!\left(\frac{\beta\omega}{4}\right) \,,
\end{aligned}
\label{eq:twopoint-q}
\end{equation}
and
\begin{equation}
\begin{aligned}
\langle x^2 \rangle_q &= \langle y^2 \rangle_q = \frac{1}{2\omega} \coth\!\left(\frac{\beta\omega}{2}\right) \,, &\qquad
\langle xy \rangle_q &= \frac{1}{2\omega \sinh(\beta\omega/2)} \,, \\
\langle\eta^2\rangle_q &= \langle\beta^2\rangle_q = \frac{\omega}{2} \coth\!\left(\frac{\beta\omega}{2}\right) \,, &\qquad
\langle\eta\beta\rangle_q &= \frac{\omega}{2 \sinh(\beta\omega/2)} \,.
\end{aligned}
\label{eq:twopoint-xy-q}
\end{equation}

The Liouville operator is invariant under the discrete parity transformation
\begin{equation}
\begin{aligned}
\hat P:\ &\hat z \to -\hat z,\ \ \hat v \to -\hat v,\ \ \hat\zeta \to -\hat\zeta,\ \ \hat\gamma \to -\hat\gamma,\\
&\hat x \to -\hat x,\ \ \hat y \to -\hat y,\ \ \hat\eta \to -\hat\eta,\ \ \hat\beta \to -\hat\beta \,.
\end{aligned}
\label{eq:parity-P}
\end{equation}
This also holds for the operators $H^{(q)}$ and $\hat E^{(cl)}$. Both the quantum and the classical thermal equilibrium are invariant under $\hat P$. (The eigenfunctions $\psi_m$ are either even or odd under $\hat P$.) As a consequence, the expectation values $\langle\hat z\rangle, \langle\hat v\rangle$ etc.\ all vanish. Both the quantum and classical thermal equilibrium are also invariant under time reversal
\begin{equation}
\hat T:\ \hat v \to -\hat v,\ \ \hat\zeta \to -\hat\zeta,\ \ \hat x \leftrightarrow \hat y,\ \ \hat\eta \leftrightarrow \hat\beta \,.
\label{eq:time-reversal-T}
\end{equation}
This implies for many combinations the vanishing of mixed correlation functions, as $\langle\hat z\hat v\rangle=0$, $\langle\hat\gamma\hat\zeta\rangle=0$. In particular, the relations
\begin{equation}
\langle z\zeta\rangle = 0 \,, \qquad \langle\gamma v\rangle = 0 \,,
\label{eq:zzeta-gammav-zero}
\end{equation}
imply both for the classical and quantum equilibrium
\begin{equation}
\langle x^2\rangle = \langle y^2\rangle = \langle z^2\rangle + \frac14 \langle\zeta^2\rangle \,, \qquad
\langle\eta^2\rangle = \langle\beta^2\rangle = \langle v^2\rangle + \frac14 \langle\gamma^2\rangle \,.
\label{eq:twopoint-PT-relations}
\end{equation}
We may use this to obtain for the classical equilibrium
\begin{equation}
\langle\eta^2\rangle_{cl} = \frac{1}{\beta} + \frac{\beta\omega^2}{16} \,, \qquad
\langle x^2\rangle_{cl} = \frac{1}{\omega^2} \langle\eta^2\rangle_{cl} \,,
\label{eq:twopoint-cl-explicit}
\end{equation}
in accordance with eqs.~\eqref{eq:Eq-cl-explicit}, \eqref{eq:Deltaq-cl-lowT}.

\section{Free classical field theory}
\label{sec:fieldtheory}

A free classical field theory is a system of uncoupled momentum modes, each described by a harmonic oscillator. We can therefore largely take over the results of the preceding section.

\subsection{Free field in momentum space}
\label{sec:field-momentum}

We discuss a free complex scalar field $\sigma(\vec x)$ and its time derivative $\pi(\vec x) = \partial_t \sigma(\vec x)$. The field equation reads
\begin{equation}
\partial_t \sigma(\vec x) = \pi(\vec x) \,, \qquad \partial_t \pi(\vec x) = -B\, \sigma(\vec x) \,,
\label{eq:field-eom}
\end{equation}
with Klein-Gordon operator ($\partial_k = \partial/\partial x_k$)
\begin{equation}
B = -\Delta + m^2 = -\partial_k \partial_k + m^2 \,.
\label{eq:KG-operator}
\end{equation}
At every time $t$ the classical probability distribution $w[\sigma,\pi]$ is a functional of the field configurations $\sigma(\vec x)$ and $\pi(\vec x)$ -- we deal with a probabilistic classical field theory. It is positive and normalized,
\begin{equation}
\int \tilde D\sigma\, \tilde D\pi\, w[\sigma,\pi] = 1 \,,
\label{eq:w-normalization}
\end{equation}
with $\int \tilde D\sigma\, \tilde D\pi$ an appropriate functional measure. Initial conditions are given at time $t$ by a probability distribution, and the time evolution of the probability distribution is governed by the Liouville equation
\begin{equation}
\partial_t w = -\hat L\, w \,,
\label{eq:field-liouville}
\end{equation}
with Liouville operator
\begin{equation}
\hat L = \int_{\vec x} \left( \pi(\vec x) \frac{\partial}{\partial \sigma(\vec x)} - B\sigma(\vec x) \frac{\partial}{\partial \pi(\vec x)} + c.c. \right) \,.
\label{eq:field-liouville-op}
\end{equation}

A Fourier transformation to momentum space is a variable transformation for the arguments of $w$, with unit Jacobian,
\begin{equation}
\sigma(\vec p) = \int_x \exp(-i\vec p\vec x)\, \sigma(\vec x) \,, \qquad
\pi(\vec p) = \int_x \exp(-i\vec p\vec x)\, \pi(\vec x) \,.
\label{eq:field-fourier}
\end{equation}
The Liouville equation becomes in momentum space
\begin{equation}
\hat L = \int_{\vec p} \left( \pi(\vec p) \frac{\partial}{\partial \sigma(\vec p)} - B\sigma(\vec p) \frac{\partial}{\partial \pi(\vec p)} + c.c. \right) \,,
\label{eq:field-liouville-momentum}
\end{equation}
where we use conventions with a dimensionless momentum integral,
\begin{equation}
\int_{\vec p} = \frac{\Omega}{(2\pi)^3} \int d^3p \,.
\label{eq:momentum-integral-def}
\end{equation}
(For a finite volume of space $\Omega$, say a torus, $\int_{\vec p}$ becomes a sum over the discrete momentum modes.) The operator $B$ is diagonal in momentum space,
\begin{equation}
B = \vec p^{\,2} + m^2 \,.
\label{eq:B-momentum}
\end{equation}

The field equation \eqref{eq:field-eom} describes the dynamics of independent harmonic oscillators for every value of $\vec p$,
\begin{equation}
\partial_t \sigma(\vec p) = \pi(\vec p) \,, \qquad \partial_t \pi(\vec p) = -\omega^2(\vec p)\, \sigma(\vec p) \,,
\label{eq:field-eom-momentum}
\end{equation}
where $\omega^2(\vec p) = \vec p^{\,2} + m^2$, and we can consider the real and imaginary parts of $\sigma(\vec p)$ and $\pi(\vec p)$ as two independent oscillators. (For a real classical scalar field one has $\sigma^*(\vec p) = \sigma(-\vec p)$, $\pi^*(\vec p) = \pi(-\vec p)$, such that only one independent oscillator remains for every $\vec p$.)

While the dynamics for the different Fourier modes decouples, the probability distribution is, in general, not a product of probability distributions for every $\vec p$-mode. There can be correlations between the different modes.

\subsection{Quantum field theory}
\label{sec:qft}

The probabilistic classical field theory for a free complex scalar field comprises as a subsystem the quantum field theory for a free complex scalar field \cite{CWQF}. We can write the Hamiltonian in field space in terms of hermitian operators,
\begin{equation}
H_L = -i\hat L = \int_{\vec p} \Big( \hat\pi^*(\vec p)\, \hat\gamma(\vec p) + (\vec p^{\,2}+m^2)\, \hat\sigma^*(\vec p)\, \hat\zeta(\vec p) + c.c. \Big) \,,
\label{eq:HL-field-hat}
\end{equation}
with
\begin{equation}
\hat\sigma(\vec p) = \sigma(\vec p) \,, \quad \hat\gamma(\vec p) = -i\frac{\partial}{\partial \sigma^*(\vec p)} \,, \qquad
\hat\pi(\vec p) = \pi(\vec p) \,, \quad \hat\zeta(\vec p) = i\frac{\partial}{\partial \pi^*(\vec p)} \,.
\label{eq:field-hat-ops}
\end{equation}
With $\hat\sigma(\vec p) = (\hat\sigma_1(\vec p)+i\hat\sigma_2(\vec p))/\sqrt2$ etc., we can write $H_L$ as a sum of one-dimensional classical harmonic oscillators \eqref{eq:HL-explicit},
\begin{equation}
H_L = \sum_j \int_{\vec p} \Big( \hat\pi_j(\vec p)\, \hat\gamma_j(\vec p) + (\vec p^{\,2}+m^2)\, \hat\sigma_j(\vec p)\, \hat\zeta_j(\vec p) \Big) \,,
\label{eq:HL-field-sum}
\end{equation}
where one substitutes
\begin{equation}
\hat z \to \sigma_j(\vec p) \,, \quad \hat v \to \pi_j(\vec p) \,, \quad
\hat\gamma \to -i\frac{\partial}{\partial \sigma_j(\vec p)} \,, \quad \hat\zeta \to i\frac{\partial}{\partial \pi_j(\vec p)} \,.
\label{eq:field-hat-substitutions}
\end{equation}

We can again perform the decomposition \eqref{eq:new-ops}, \eqref{eq:HL-decomp},
\begin{equation}
\begin{aligned}
\hat\varphi(\vec p) &= \hat\sigma(\vec p) + \tfrac12 \hat\zeta(\vec p) \,, & \hat\chi(\vec p) &= \hat\sigma(\vec p) - \tfrac12 \hat\zeta(\vec p) \,, \\
\hat\eta(\vec p) &= \hat\pi(\vec p) + \tfrac12 \hat\gamma(\vec p) \,, & \hat\beta(\vec p) &= \hat\pi(\vec p) - \tfrac12 \hat\gamma(\vec p) \,,
\end{aligned}
\label{eq:field-xchi-ops}
\end{equation}
and $H_L = H^{(q)} - H^{(m)}$,
\begin{equation}
\begin{aligned}
H^{(q)} &= \int_{\vec p} \Big( \hat\eta^*(\vec p)\,\hat\eta(\vec p) + \omega^2(\vec p)\, \hat\varphi^*(\vec p)\, \hat\varphi(\vec p) \Big) \,, \\
H^{(m)} &= \int_{\vec p} \Big( \hat\beta^*(\vec p)\,\hat\beta(\vec p) + \omega^2(\vec p)\, \hat\chi^*(\vec p)\, \hat\chi(\vec p) \Big) \,.
\end{aligned}
\label{eq:HqHm-field}
\end{equation}
with
\begin{equation}
\omega^2(\vec p) = m^2 + \vec p^{\,2} \,.
\label{eq:omega2-def}
\end{equation}

The operators $\hat x$ and $\hat y$ are now replaced by field operators $\hat\varphi(\vec p)$ and $\hat\chi(\vec p)$, and similarly for $\hat\eta$ and $\hat\beta$. The non-vanishing commutators for field operators are given by
\begin{equation}
[\hat\varphi_i(\vec p), \hat\eta_j(\vec p\,')] = i\,\delta_{ij}\,\delta(\vec p-\vec p\,') \,, \qquad
[\hat\chi_i(\vec p), \hat\beta_j(\vec p\,')] = -i\,\delta_{ij}\,\delta(\vec p-\vec p\,') \,.
\label{eq:field-commutators}
\end{equation}
The subset of operators $\hat\varphi(\vec p)$ and $\hat\eta(\vec p)$ constitutes a quantum field theory with dynamics determined by the free Hamiltonian $H^{(q)}$.

We define the classical wave function as the square root of the probability distribution,
\begin{equation}
w[\sigma,\pi] = q^2[\sigma,\pi] \,.
\label{eq:w-classical-field}
\end{equation}
Performing a functional Fourier transform of the wave function,
\begin{equation}
\tilde\Psi[\sigma,\zeta] = \int \tilde D\pi\, \exp\Big\{ i\sum_j \int_{\vec p} \zeta_j(\vec p)\, \pi_j(\vec p) \Big\}\, q(\sigma,\pi) \,,
\label{eq:psi-tilde-field-sigma-phi}
\end{equation}
one defines $\tilde\Psi[\varphi,\chi]$ by the linear combination analogous to \eqref{eq:field-xchi-ops}. The functional integral $\int\tilde D\pi$ is a product of integrals over $\pi_j(\vec p)$.

\subsection{Quantum thermal equilibrium}
\label{sec:qte-field}

For the quantum thermal equilibrium the classical wave function $\tilde\Psi[\varphi,\chi]$ is a product of wave functions for different $\vec p$,
\begin{equation}
\tilde\Psi[\varphi,\chi] = \prod_j \prod_{\vec p} \tilde\Psi_{j,\vec p}\big(\varphi_j(\vec p), \chi_j(\vec p)\big) \,,
\label{eq:psi-tilde-field-xchi}
\end{equation}
with $\tilde\Psi_{j,\vec p}$ the analogue of eq.~\eqref{eq:psi-thermal-xy},
\begin{equation}
\tilde\Psi_{j,\vec p} = Z_{\vec p}^{-1/2} \sum_m \exp\Big\{-\tfrac12 \beta\omega(\vec p)(m+\tfrac12)\Big\}\, \psi_m\big(\varphi_j(\vec p)\big)\, \psi_m^*\big(\chi_j(\vec p)\big) \,,
\label{eq:psi-thermal-mode}
\end{equation}
and
\begin{equation}
Z_{\vec p} = \sum_m \exp\big(-\beta\omega(\vec p)(m+\tfrac12)\big) \,, \qquad \omega(\vec p) = \sqrt{\vec p^{\,2}+m^2} \,.
\label{eq:Zp-thermal}
\end{equation}
In turn, also $q[\sigma,\pi]$ and $w[\sigma,\pi]$ factorize into products for the different momentum modes, and this also holds for the quantum density matrix $\rho[\varphi,\varphi']$.

In consequence, the expectation value of the quantum energy,
\begin{equation}
\langle E^{(q)} \rangle_q = \tilde\Psi^\dagger H^{(q)} \tilde\Psi \,,
\label{eq:Eq-field-def}
\end{equation}
becomes simply a sum of the quantum energies \eqref{eq:Hq-eigen} for the individual momentum modes,
\begin{equation}
\begin{aligned}
\langle E^{(q)} \rangle_q &= \int_{\vec p} \omega(\vec p) \coth\!\left(\frac{\beta\omega(\vec p)}{2}\right)
= 2\int_{\vec p} \omega(\vec p) \big(n(\vec p)+\tfrac12\big) \\
&= E_0 + \int_{\vec p} \sqrt{\vec p^{\,2}+m^2}\, \left( \coth\!\left(\frac{\beta\sqrt{\vec p^{\,2}+m^2}}{2}\right) - 1 \right) \,,
\end{aligned}
\label{eq:Eq-field-explicit}
\end{equation}
Here $n(\vec p)$ are the mean occupation numbers,
\begin{equation}
n(\vec p) = \frac{1}{\exp(\beta\omega(\vec p))-1} \,.
\label{eq:n-occupation}
\end{equation}
The ground state energy,
\begin{equation}
E_0 = \int_{\vec p} \omega(\vec p) = \frac{\Omega}{(2\pi)^3} \int d^3p\, \sqrt{p^2+m^2} \,,
\label{eq:E0-ground}
\end{equation}
is independent of $\beta$. This constant diverges if the ultraviolet cutoff for the momentum integration is sent to infinity. Such a divergent constant does not matter for observations.

The temperature dependent part is given by the second term in eq.~\eqref{eq:Eq-field-explicit}. It is finite,
\begin{equation}
\langle E^{(q)} \rangle_q - E_0 = \frac{\Omega}{2\pi^2} \int_m^\infty d\omega\, \omega^2 \sqrt{\omega^2-m^2}\, \left( \coth\!\left(\frac{\beta\omega}{2}\right) - 1 \right) \,,
\label{eq:Eq-field-finite}
\end{equation}
due to the exponential suppression $\coth(x)-1 = 2\exp(-2x)$ for $x\to\infty$.
Subtracting the irrelevant ground state energy one finds for the frequency distribution of the quantum energy
\begin{equation}
\langle E^{(q)} \rangle - E_0 = 2\int_{\vec p} \omega(\vec p)\, n(\vec p) \,,
\label{eq:Eq-freq-distribution}
\end{equation}
The temperature dependent mean occupation number $n(\vec p)$ describes the spectrum for the thermal equilibrium state of a free quantum field theory. In the massless limit $m\to0$ this is precisely the Planck spectrum.

The situation is similar for the expectation value of the classical energy in the quantum thermal state, $\langle E^{(cl)} \rangle_q$. The divergent constant is smaller by a factor two, and the temperature dependent part remains finite. This result obtains directly by a sum over the different momentum modes of the expression for the harmonic oscillator in the preceding section.

The free energy,
\begin{equation}
W(\beta) = \ln Z(\beta) = \sum_j \int_{\vec p} \ln Z_{\vec p}(\beta)
= W_0 - 2\int_{\vec p} \ln\big(1-\exp(-\beta\omega(\vec p))\big) \,,
\label{eq:W-field-free-energy}
\end{equation}
involves a divergent part
\begin{equation}
W_0 = -\beta \int_{\vec p} \omega(\vec p) = -\beta E_0 \,,
\label{eq:W0-divergent}
\end{equation}
while the temperature dependent part remains finite. The first derivative with respect to $\beta$ yields the expectation value of the quantum energy according to
\begin{equation}
\langle E^{(q)} \rangle_q = -\frac{\partial}{\partial\beta} W(\beta) \,.
\label{eq:Eq-field-from-W}
\end{equation}
All higher $\beta$-derivatives, as the specific heat, remain finite. All this is precisely what one knows from quantum field theory.

The (equal-time) two-point function for the fluctuating field $\varphi(\vec p)$ reads for the quantum equilibrium
\begin{equation}
\langle \varphi_i(\vec p)\, \varphi_j(\vec p\,') \rangle_q = G(\vec p)\, \delta_{ij}\, \delta(\vec p-\vec p\,') \,,
\label{eq:twopoint-field-q}
\end{equation}
where $G(\vec p)$ corresponds to $\langle x^2 \rangle_q$ in eq.~\eqref{eq:twopoint-xy-q},
\begin{equation}
G(\vec p) = \frac{1}{2\omega(\vec p)} \coth\!\left(\frac{\beta\omega(\vec p)}{2}\right) = \frac{1}{\omega(\vec p)}\left(n(\vec p)+\tfrac12\right) \,.
\label{eq:G-field-explicit}
\end{equation}
For the time-derivative operator one finds
\begin{equation}
\langle \eta_i(\vec p)\, \eta_j(\vec p\,') \rangle = \omega^2(\vec p)\, \langle \varphi_i(\vec p)\, \varphi_j(\vec p\,') \rangle \,.
\label{eq:eta-twopoint-field}
\end{equation}
The vacuum propagator obtains in the limit $\beta\to\infty$,
\begin{equation}
G_0(\vec p) = \frac{1}{2\omega(\vec p)} = \frac{1}{2\sqrt{\vec p^{\,2}+m^2}} \,.
\label{eq:G0-vacuum}
\end{equation}
This is the standard (equal-time) propagator for the quantum field theory of a free massive complex scalar field.

\subsection{Classical thermal equilibrium}
\label{sec:cte-field}

The classical energy for the classical field theory is given by
\begin{equation}
E^{(cl)} = \frac12 \sum_j \int_{\vec p} \Big( \pi_j(\vec p)^2 + \omega^2(\vec p)\, \sigma_j(\vec p)^2 \Big) \,,
\label{eq:Ecl-field-def}
\end{equation}
with a corresponding operator expression for $\hat E^{(cl)}$. The classical thermal equilibrium is characterized according to the Boltzmann probability distribution by the wave function
\begin{equation}
q[\sigma,\pi] = Z_{cl}^{-1/2} \exp\!\left(-\frac{\beta}{2} E^{(cl)}[\sigma,\pi]\right) \,.
\label{eq:q-classical-field}
\end{equation}
In momentum space the Gaussian distribution factorizes into contributions from different momentum modes,
\begin{equation}
q[\sigma,\pi] = \prod_j \prod_{\vec p} q_{j,\vec p}\big(\sigma_j(\vec p), \pi_j(\vec p)\big) \,.
\label{eq:q-classical-field-factorized}
\end{equation}

We can take over the results of the preceding section,
\begin{equation}
W_{cl} = \ln Z_{cl}(\beta) = -2\int_{\vec p} \ln\big(\beta\omega(\vec p)\big) \,,
\label{eq:Wcl-field}
\end{equation}
and
\begin{equation}
\langle E^{(cl)} \rangle_{cl} = -\frac{\partial}{\partial\beta} \ln Z_{cl}(\beta) = 2\int_{\vec p} \frac{1}{\beta} \,.
\label{eq:Ecl-field-cl}
\end{equation}
The momentum integrals \eqref{eq:Wcl-field}, \eqref{eq:Ecl-field-cl} diverge if the UV-momentum cutoff is taken to infinity. This is the Rayleigh-Jeans divergence. Also the energy in a given frequency interval and higher $\beta$-derivatives of $W_{cl}(\beta)$ diverge. The quantum energy in the classical equilibrium diverges even stronger than the classical energy, according to eq.~\eqref{eq:Eq-cl-explicit},
\begin{equation}
\langle E^{(q)} \rangle_{cl} = \langle E^{(cl)} \rangle_{cl} + \frac{\beta}{8} \int_{\vec p} \omega^2(\vec p) \,.
\label{eq:Eq-field-cl}
\end{equation}

According to eq.~\eqref{eq:twopoint-cl} the two-point functions for the classical fields obey in the classical equilibrium
\begin{equation}
\langle \sigma_i(\vec p)\, \sigma_j(\vec p\,') \rangle_{cl} = \delta_{ij}\, \delta(\vec p-\vec p\,')\, \frac{1}{\beta\omega^2(\vec p)} \,, \qquad
\langle \pi_i(\vec p)\, \pi_j(\vec p\,') \rangle_{cl} = \delta_{ij}\, \delta(\vec p-\vec p\,')\, \frac{1}{\beta} \,.
\label{eq:twopoint-field-cl}
\end{equation}
These correlation functions vanish for the classical ground state, $\beta\to\infty$. The classical ground state is a sharp field configuration with $\sigma_j(\vec p)=0$, $\pi_j(\vec p)=0$,
\begin{equation}
w_{cl,0} = \prod_j \prod_{\vec p} \delta\big(\sigma_j(\vec p)\big)\, \delta\big(\pi_j(\vec p)\big) \,.
\label{eq:w0-classical-field}
\end{equation}
Due to $T$-invariance the mixed correlation function vanishes both for the classical and quantum equilibrium,
\begin{equation}
\langle \sigma_i(\vec p)\, \pi_j(\vec p\,') \rangle = 0 \,.
\label{eq:mixed-twopoint-zero-field}
\end{equation}

\subsection{Comparison at fixed classical energy}
\label{sec:comparison-fixedE}

The quantum and classical equilibrium states lead to different momentum distributions for the fields $\sigma$ and $\pi$. We define the mean classical energy per mode,
\begin{equation}
\tilde T(\vec p) = \frac12 \langle \pi_j(\vec p)\, \pi_j(\vec p) \rangle = \frac{\omega^2(\vec p)}{2} \langle \sigma_j(\vec p)\, \sigma_j(\vec p) \rangle \,.
\label{eq:Ttilde-def}
\end{equation}
The mean value of the classical energy is given by the momentum integral
\begin{equation}
\langle E^{(cl)} \rangle = 2 \int_{\vec p} \tilde T(\vec p) \,.
\label{eq:Ecl-mode-integral}
\end{equation}
The classical energy is a conserved quantity. If we want to compare how a given classical energy is distributed over the different momentum modes, we need to make the comparison for a fixed value of the classical energy,
\begin{equation}
\langle E^{(cl)} \rangle_{cl} = \langle E^{(cl)} \rangle_q = \bar E \,.
\label{eq:fixed-E-condition}
\end{equation}
A well-defined value of $\bar E$ requires an ultraviolet cutoff for the momentum integral ($\tilde p=\sqrt{\vec p^{\,2}}$),
\begin{equation}
\int_{\vec p} = \frac{\Omega}{(2\pi)^3} \int_0^\Lambda 4\pi \tilde p^2\, d\tilde p \,.
\label{eq:momentum-cutoff-integral}
\end{equation}
For a given $T$ the mean values of the classical energy differ for the quantum and thermal equilibrium.

\begin{figure}[t]
\centering
\includegraphics[width=0.62\textwidth]{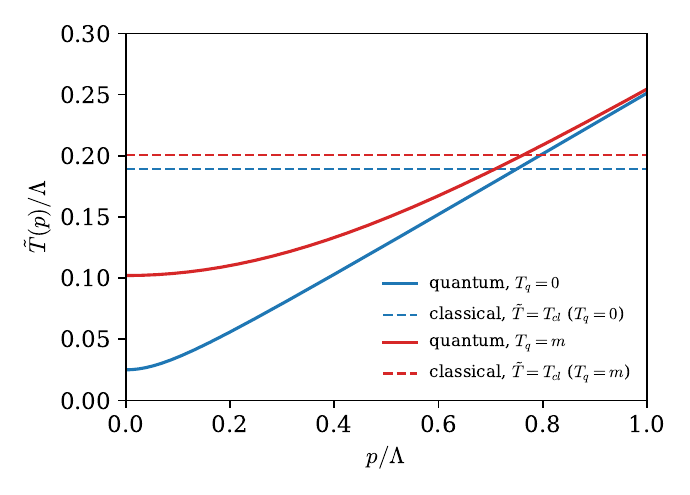}
\caption{Mode temperature $\tilde T(p)$, in units of the cutoff $\Lambda$, for the quantum equilibrium (solid lines, eq.~\eqref{eq:Ttilde-q}) and for the classical equilibrium with the same mean classical energy $\langle E^{(cl)} \rangle$ (dashed horizontal lines, $\tilde T = T_{cl}$). We take $m=\Lambda/10$ and $T_q=0$ (blue) or $T_q=m$ (red).}
\label{fig:Ttilde}
\end{figure}

For a comparison at fixed classical energy we therefore use different temperatures $T_q$ and $T_{cl}$ for the quantum and classical equilibrium. For the classical equilibrium one has
\begin{equation}
\tilde T_{cl}(\vec p) = T_{cl} \,,
\label{eq:Ttilde-cl}
\end{equation}
and
\begin{equation}
\bar E = \frac{\Omega\Lambda^3}{3\pi^2}\, T_{cl} \,.
\label{eq:Ebar-cl}
\end{equation}
For the quantum equilibrium the temperature $T_q$ as a function of $\bar E$ is determined by
\begin{equation}
\bar E = 2\int_{\vec p} \frac{\omega(\vec p)}{4} \coth\!\left(\frac{\omega(\vec p)}{4T_q}\right) \,,
\label{eq:Ebar-quantum-Tq}
\end{equation}
such that the temperatures $T_q$ and $T_{cl}$ are related by
\begin{equation}
\int_{\vec p} \left( \frac{\omega(\vec p)}{4} \coth\!\left(\frac{\omega(\vec p)}{4T_q}\right) - T_{cl} \right) = 0 \,.
\label{eq:Tq-Tcl-relation}
\end{equation}
The momentum distribution for the quantum equilibrium obeys
\begin{equation}
\tilde T_q(\vec p) = \frac{\omega(\vec p)}{4} \coth\!\left(\frac{\omega(\vec p)}{4T_q}\right) \,,
\label{eq:Ttilde-q}
\end{equation}
which can now be compared with the flat distribution \eqref{eq:Ttilde-cl} of the classical equilibrium.
In fig.~\ref{fig:Ttilde} we compare the ``mode temperature'' $\tilde T(p)$ for the classical and quantum equilibrium at fixed mean classical energy $\langle E^{(cl)} \rangle$. We take $m=\Lambda/10$ and show the dependence on $p$ for two values of $T_q$, $T_q=0$ and $T_q=m$.

For the quantum equilibrium one may define a ``mean frequency'' $\bar\omega$ by
\begin{equation}
\bar E = 2\int_{\vec p} \frac{\bar\omega}{4} \coth\!\left(\frac{\bar\omega}{4T_q}\right) \,.
\label{eq:Ebar-meanfreq}
\end{equation}
The mean frequency is a function of $\bar E$, $\Lambda$, and $T_q$. In turn, the relation between $T_q$ and $T_{cl}$ depends on $\bar\omega$ through
\begin{equation}
\tanh\!\left(\frac{\bar\omega}{4T_q}\right) = \frac{\bar\omega}{4T_{cl}} \,,
\label{eq:tanh-meanfreq-relation}
\end{equation}
or
\begin{equation}
T_q = \frac{\bar\omega}{2\ln\!\left(\dfrac{4T_{cl}+\bar\omega}{4T_{cl}-\bar\omega}\right)} \,.
\label{eq:Tq-meanfreq-explicit}
\end{equation}
For a given $\bar\omega$ the ratio between the momentum distributions in the quantum and classical equilibrium is given by
\begin{equation}
\frac{\tilde T_q(\vec p)}{\tilde T_{cl}(\vec p)} = \frac{\omega(\vec p)\, \coth\!\big(\omega(\vec p)/4T_q\big)}{\bar\omega\, \coth\!\big(\bar\omega/4T_q\big)} \,.
\label{eq:ratio-Tq-Tcl-meanfreq}
\end{equation}
Thus $\bar\omega$ denotes the frequency for which $\tilde T_q(\vec p)$ equals $\tilde T_{cl}$.

The quantum equilibrium is characterized by the mass scales $T_q$, $m$, and $\Lambda$. If we choose $\Lambda \gg m, T_q$, the classical energy \eqref{eq:Ebar-cl}, \eqref{eq:Ebar-quantum-Tq} is dominated by the modes with $p^2 \approx \Lambda^2$,
\begin{equation}
\bar E \approx \frac12 \int_{\vec p} \sqrt{m^2+p^2} \approx \frac{\Lambda^4 \Omega}{16\pi^2} \,.
\label{eq:Ebar-cutoff-dominated}
\end{equation}
Comparison with eq.~\eqref{eq:Ebar-meanfreq} yields in this case
\begin{equation}
\bar\omega \approx \frac{3\Lambda}{4} \,.
\label{eq:meanfreq-cutoff}
\end{equation}
From eq.~\eqref{eq:tanh-meanfreq-relation} one concludes that the corresponding classical temperature $T_{cl}$ for the classical equilibrium with the same $\langle E^{(cl)} \rangle$ has to be near the cutoff,
\begin{equation}
T_{cl} \approx \frac{\bar\omega}{4} \approx \frac{3\Lambda}{16} \,.
\label{eq:Tcl-cutoff}
\end{equation}
For $T_q \ll \Lambda$ one finds $T_{cl}$ much larger than $T_q$.

For $\bar\omega \gg 4T_q$ the ratio of momentum distributions \eqref{eq:ratio-Tq-Tcl-meanfreq} simplifies to
\begin{equation}
\frac{\tilde T_q(\vec p)}{\tilde T_{cl}(\vec p)} = \frac{4\sqrt{m^2+\vec p^{\,2}}}{3\Lambda} \coth\!\left(\frac{\sqrt{m^2+\vec p^{\,2}}}{4T_q}\right) \,.
\label{eq:ratio-cutoff-limit}
\end{equation}
For $\omega(\vec p) \ll \Lambda$ the quantum spectrum is suppressed as compared to the classical spectrum. This differs strongly from a comparison where one would simply replace $T_{cl}$ by $T_q$. For $\omega(\vec p) \gtrsim T_q$ the suppression factor is of order $\omega(\vec p)/\Lambda$, while for $\omega(\vec p) \lesssim T_q$ it is of order $T_q/\Lambda$.

This extends to a comparison of the correlation functions for the classical field $\sigma$ at equal $\langle E^{(cl)} \rangle$. With
\begin{equation}
\langle \sigma_i(\vec p)\, \sigma_j(\vec p\,') \rangle = H(\vec p)\, \delta_{ij}\, \delta(\vec p - \vec p\,') \,, \qquad
H(\vec p) = \frac{2\tilde T(\vec p)}{\omega^2(\vec p)} \,,
\label{eq:H-def}
\end{equation}
the ratio is again given by eq.~\eqref{eq:ratio-Tq-Tcl-meanfreq},
\begin{equation}
\frac{H_q(\vec p)}{H_{cl}(\vec p)} = \frac{\tilde T_q(\vec p)}{\tilde T_{cl}(\vec p)} \,.
\label{eq:H-ratio}
\end{equation}
Towards small $\omega(\vec p)$ the enhancement $\propto \omega(\vec p)^{-2}$ for the classical equilibrium is stronger than the enhancement $\propto \omega(\vec p)^{-1}$ for the quantum equilibrium.

A temperature $T_{cl} \ll \Lambda$ is not compatible with a quantum equilibrium with the same $\langle E^{(cl)} \rangle$. The value \eqref{eq:Ebar-cutoff-dominated} is a lower bound for $\langle E^{(cl)} \rangle_q$, which exceeds the value \eqref{eq:Ebar-cl} for $\langle E^{(cl)} \rangle_{cl}$. For arbitrary $m$ and $T_q$ one has
\begin{equation}
\langle E^{(cl)} \rangle_q \geq \frac{\Lambda^4 \Omega}{16\pi^2} \,,
\label{eq:Eq-cl-lower-bound}
\end{equation}
such that eq.~\eqref{eq:Ebar-cl} results in a lower bound for $T_{cl}$,
\begin{equation}
T_{cl} > \frac{3\Lambda}{16} \,.
\label{eq:Tcl-lower-bound}
\end{equation}

One can, of course, find a classical thermal equilibrium distribution with $T_{cl} \ll \Lambda$. There exists simply no quantum thermal equilibrium distribution with the same mean classical energy. The reason is that the mean classical energy in the quantum equilibrium has to exceed the one for the ground state. These considerations matter if one investigates the time evolution starting with probability distributions different from the equilibrium distributions, possibly adding small interactions. A state close to the quantum equilibrium is reached in the long time limit only if the initial classical energy $\langle E^{(cl)} \rangle$ is large enough. The conserved classical energy includes for the quantum equilibrium the ground state energy.

\subsection{Energy dispersion}
\label{sec:energy-dispersion}

Let us compare the fluctuations in the classical energy for the quantum and classical thermal equilibrium states,
\begin{equation}
\begin{aligned}
(\Delta E^{(cl)})^2 &= \langle E^{(cl)2} \rangle - \langle E^{(cl)} \rangle^2 \\
&= \int \tilde D\sigma\, \tilde D\pi\; q(\sigma,\pi)\, (\hat E^{(cl)})^2\, q(\sigma,\pi) - \langle E^{(cl)} \rangle^2 \,,
\end{aligned}
\label{eq:DeltaEcl-def}
\end{equation}
with $\hat E^{(cl)}$ given by eq.~\eqref{eq:Ecl-field-def}. For the classical equilibrium state one finds
\begin{equation}
(\Delta E^{(cl)})^2_{cl} = \frac{\partial^2}{\partial\beta^2} \ln Z_{cl}
= -2\int_{\vec p} \frac{\partial^2}{\partial\beta^2} \ln\big(\beta\omega(\vec p)\big)
= 2\int_{\vec p} \frac{1}{\beta^2} = \frac{\Omega\Lambda^3}{3\pi^2\beta^2} \,.
\label{eq:DeltaEcl-explicit}
\end{equation}
We can identify $2\int_{\vec p} 1$ with the total number of degrees of freedom $N$. This yields
\begin{equation}
\frac{(\Delta E^{(cl)})^2_{cl}}{\langle E^{(cl)} \rangle_{cl}^2} = \frac{1}{N} \,,
\label{eq:DeltaEcl-ratio-N}
\end{equation}
as characteristic for a Gaussian distribution.

In comparison, one may evaluate the dispersion of the classical energy for the quantum thermal equilibrium. This becomes particularly simple for the quantum ground state, $T_q=0$, for which the classical probability distribution is a Gaussian, cf. eq.~\eqref{eq:w-classical-field},
\begin{equation}
w = \prod_j \prod_{\vec p} \left[ 4\exp\left\{ -\frac{4}{\omega(\vec p)} \left( \frac{\pi_j(\vec p)^2}{2} + \frac{\omega^2(\vec p)}{2}\sigma_j(\vec p)^2 \right) \right\} \right] \,.
\label{eq:w-groundstate-field}
\end{equation}
One finds
\begin{equation}
(\Delta E^{(cl)})^2_q = \langle E^{(cl)2} \rangle_q - \langle E^{(cl)} \rangle_q^2 = 2\int_{\vec p} \Delta E_p^2 \,,
\label{eq:DeltaEcl-q-def}
\end{equation}
where $\Delta E_p^2$ denotes the energy dispersion for a given momentum mode in the quantum ground state,
\begin{equation}
\Delta E_p^2 = \frac{\omega^2(\vec p)}{16} = \frac{m^2+\vec p^{\,2}}{16} \,.
\label{eq:DeltaEp2-def}
\end{equation}
Comparing with the classical thermal equilibrium \eqref{eq:DeltaEcl-explicit}, one replaces $T$ in the momentum integral by $\omega(\vec p)/4$. The integral yields
\begin{equation}
(\Delta E^{(cl)})^2_q = \frac{\Omega}{80\pi^2} \left( \Lambda^5 + \frac53 m^2 \Lambda^3 \right) \,,
\label{eq:DeltaEcl-q-explicit}
\end{equation}
and one finds for the ratio
\begin{equation}
\frac{(\Delta E^{(cl)})^2_q}{\langle E^{(cl)} \rangle_q^2} = \frac{f}{N} \,,
\label{eq:DeltaEcl-q-ratio}
\end{equation}
with
\begin{equation}
f = \frac{\int_{\vec p} \omega^2(\vec p)}{\int_{\vec p} \bar\omega^2} = \frac{3\Lambda^2 + 5m^2}{5\bar\omega^2} \,.
\label{eq:f-def}
\end{equation}
For $m^2 \ll \Lambda^2$ we approximate
\begin{equation}
\bar\omega = \frac34 \Lambda \left(1+\frac{m^2}{\Lambda^2}\right) \,, \qquad f = \frac{16}{15}\left(1-\frac{m^2}{3\Lambda^2}\right) \,.
\label{eq:f-approx}
\end{equation}

Comparing eq.~\eqref{eq:DeltaEcl-q-explicit} with eq.~\eqref{eq:DeltaEcl-explicit} one observes for small $T_{cl}\ll\Lambda$ a strong enhancement of the quantum dispersion compared to the classical dispersion. If we compare at fixed $\langle E^{(cl)} \rangle_q = \langle E^{(cl)} \rangle_{cl}$, however, the ratio $(\Delta E)^2_q/(\Delta E)^2_{cl}$ is given by $f$. This is rather close to one, with a modest enhancement of the quantum dispersion. We can extend this comparison to higher order correlations for the classical energy. They are all conserved by the time evolution.

\section{Quantum-classical connection}
\label{sec:qc-connection}

For a free classical field theory we have computed the expectation value of the quantum energy in the classical thermal equilibrium state, and the classical energy in the quantum thermal equilibrium state. In this section we discuss the relation between quantum and classical concepts in more detail. We also generalize to a wider class of classical field theories, allowing for inhomogeneous dynamics and interactions. Beyond the equilibrium states we describe the time evolution of the density matrix and the correlation functions. Our starting point is always a classical field theory. We do not discuss the classical approximation to a quantum field theory, which is a wide and interesting subject by its own.

\subsection{Classical probability distribution, quantum density matrix and wave function}

For every classical probability distribution one can construct a quantum density matrix, and for every quantum density matrix one can compute an associated classical probability distribution. This relation is general and does not depend on the particular dynamics. By classical probability distribution we understand a positive and normalized probability distribution. This should not be confounded with the Wigner function, which obtains from the quantum density matrix by a suitable Fourier transform. The Wigner function looks similar to a probability distribution. Its weights are, however, not positive semidefinite.

The first step towards a construction of a quantum density matrix $\rho(\varphi,\varphi')$ from a classical probability distribution $w(\sigma,\pi)$ casts the classical probabilistic information into the form of the classical wave function $q(\sigma,\pi)$, $w(\sigma,\pi) = q^2(\sigma,\pi)$. From $q(\sigma,\pi)$ one computes $\tilde\Psi(\sigma,\zeta)$ by a (functional) Fourier transform, and $\tilde\Psi(\varphi,\chi)$ by a linear variable transformation. The complex function $\tilde\Psi(\varphi,\chi)$ can be viewed as a hermitian matrix,
\begin{equation}
\tilde\Psi^\dagger = \tilde\Psi \,, \qquad \tilde\Psi^*(\chi,\varphi) = \tilde\Psi(\varphi,\chi) \,.
\label{eq:qc-psi-hermitian}
\end{equation}
The quantum density matrix obtains as the square of this matrix,
\begin{equation}
\rho = \tilde\Psi^2 \,, \qquad \rho(\varphi,\varphi') = \int \tilde D\chi\; \tilde\Psi(\varphi,\chi)\, \tilde\Psi(\chi,\varphi') \,.
\label{eq:qc-rho-def}
\end{equation}
The normalization of $\tilde\Psi$ implies the normalization of $\rho$,
\begin{equation}
\mathrm{tr}\,\rho = \int \tilde D\varphi\; \rho(\varphi,\varphi) = \int \tilde D\varphi\, \tilde D\chi\; \tilde\Psi^*(\varphi,\chi)\, \tilde\Psi(\varphi,\chi) = 1 \,.
\label{eq:qc-trace-norm}
\end{equation}
Furthermore, the hermiticity of $\tilde\Psi$ implies that $\rho$ is a hermitian positive matrix,
\begin{equation}
\rho = \tilde\Psi^\dagger \tilde\Psi \,.
\label{eq:qc-rho-posdef}
\end{equation}
The hermiticity of $\tilde\Psi$, eq.~\eqref{eq:qc-psi-hermitian}, is crucial for this construction. In turn, this follows from the fact that $q(\sigma,\pi)$ is a real wave function.

In the other direction, one can find for every positive hermitian matrix $\rho$ a complex function $\tilde\Psi$ as the root in the matrix sense, eq.~\eqref{eq:qc-rho-def}. From $\tilde\Psi(\varphi,\chi)$ one computes $\tilde\Psi(\sigma,\zeta)$ by a variable transformation, and $q(\sigma,\pi)$ by inverting the Fourier transform. In turn, this defines $w(\sigma,\pi) = q^2(\sigma,\pi)$.

From the conceptual point of view the simplest object is the real wave function $q(\sigma,\pi)$. It is directly related to the complex wave function $\tilde\Psi(\varphi,\chi)$ which is often the simplest object for practical computations. The complex wave function $\tilde\Psi$ allows the computation of expectation values for both classical and statistical observables from the associated operators
\begin{equation}
\langle A \rangle = \int_{\varphi,\chi,\varphi',\chi'} \tilde\Psi^*(\varphi,\chi)\, \hat A(\varphi,\chi;\varphi',\chi')\, \tilde\Psi(\varphi',\chi') \,.
\label{eq:qc-expectation}
\end{equation}
This includes operators $\hat A$ which may involve derivatives with respect to $\varphi'$ or $\chi'$.

For observables represented by operators of the type
\begin{equation}
\hat A(\varphi,\chi;\varphi',\chi') = \hat A_q(\varphi,\varphi')\, \delta(\chi,\chi') \,,
\label{eq:cq-A-factorizing}
\end{equation}
eq.~\eqref{eq:qc-expectation} implies the standard quantum rule
\begin{equation}
\langle A \rangle = \operatorname{tr}(\rho\, \hat A) \,.
\label{eq:cq-trace-rule}
\end{equation}
This shows that the quantum density matrix can be interpreted as a subsystem for observables involving only $\varphi$.

The relation \eqref{eq:qc-expectation} allows for a basis transformation of $\tilde\Psi$. In one direction this is employed for the computation of $q(\sigma,\pi)$ and $w(\sigma,\pi)$ (the classical probability distribution). In the other direction the quantum density matrix obtains by the squared matrix \eqref{eq:qc-psi-hermitian}, \eqref{eq:qc-rho-def} in an arbitrary basis. The computation of $\tilde\Psi$ from $\rho$ may be cumbersome for general $\rho$, but is guaranteed to exist by the possibility to diagonalize $\rho$. Both the transformations $w\to q$ and $\rho\to\tilde\Psi$ involve sign ambiguities. These will not matter for most practical purposes -- the signs being largely fixed by continuity requirements.

\subsection{Classical observables from quantum density matrix}

The probabilistic information in the classical probability distribution $w(\sigma,\pi)$ and the quantum density matrix $\rho(\varphi,\varphi')$ is essentially the same, both being encoded in the wave function $\tilde\Psi(\varphi,\chi)$. It is for this reason that we have been able to compute the expectation value of the classical energy from the quantum density matrix of the quantum thermal equilibrium, and the mean quantum energy from the classical probability distribution of the classical thermal equilibrium.

For the discussion of ``quantum observables'' we have extended the set of possible observables by including statistical observables. These include observables which are represented by operators involving derivatives with respect to $\sigma$ and $\pi$. The concept of the wave function $q(\sigma,\pi)$ is crucial for this purpose. Statistical observables are typically represented by operators acting linearly on the wave function. There is no linear representation on the basis of the probability distribution. For example, the ``roughness observable'' $R$ associated to the operator $\hat R = -\partial_\sigma^2$ finds only a non-linear expression in terms of $w$,
\begin{equation}
\langle R \rangle = \int_{\sigma,\pi} \Big(\partial_\sigma \sqrt{w(\sigma,\pi)}\Big)^2 = \frac14 \int_{\sigma,\pi} \frac{\partial_\sigma w\, \partial_\sigma w}{w} \,.
\label{eq:qc-roughness}
\end{equation}

The computation of classical observables $B(\sigma,\pi)$ from a quantum density matrix $\rho(\varphi,\varphi')$ may be more surprising. Classical observables are represented by operators $B(\sigma,\pi)$ which all commute among each other. They can therefore all have simultaneously sharp values. Such a large class of observables with possibly sharp values is not expected in a quantum theory. The reason why a computation of expectation values for classical observables from a quantum density matrix is possible is related to the fact that they cannot be extracted by a linear operation on $\rho$. Indeed, a general observable can be computed from $\tilde\Psi = \rho^{1/2}$ by a linear operator $\hat A$ in eq.~\eqref{eq:qc-expectation}. Only for operators independent of $\chi$,
\begin{equation}
\hat A(\varphi,\chi;\varphi',\chi') = \hat C(\varphi,\varphi')\, \delta(\chi-\chi') \,,
\label{eq:qc-A-indep-chi}
\end{equation}
this translates to the quantum rule in terms of the density matrix,
\begin{equation}
\langle A \rangle = \int_{\varphi,\varphi',\chi} \hat C(\varphi,\varphi')\, \tilde\Psi(\varphi',\chi)\, \tilde\Psi^*(\varphi,\chi)
= \int_{\varphi,\varphi'} \hat C(\varphi,\varphi')\, \rho(\varphi',\varphi) = \mathrm{tr}(\hat C \rho) \,.
\label{eq:qc-A-trace}
\end{equation}
(This holds similarly for operators independent of $\varphi$.) The operators for classical observables are not of the type \eqref{eq:qc-A-indep-chi}. The expectation values of classical observables require to take the root $\tilde\Psi = \rho^{1/2}$. They are quadratic in $\rho^{1/2}$. This is what we have done for the computation of $\langle E^{(cl)} \rangle$ in the quantum equilibrium.

In short, $w(\sigma,\pi)$ gives direct access to classical observables. For the computation of statistical observables one needs the root $q(\sigma,\pi)$ of this function. On the other hand, $\rho(\varphi,\varphi')$ gives direct access to ``quantum observables'' characterized by the property \eqref{eq:qc-A-indep-chi}. The computation of more general observables, which include the classical observables, necessitates to take the matrix root $\tilde\Psi(\varphi,\chi)$.

\subsection{Pure (and mixed) quantum states}

Pure quantum states obey
\begin{equation}
\rho^2 = \rho \,, \qquad \tilde\Psi^2 = \pm\tilde\Psi \,, \qquad \rho = \pm\tilde\Psi \,.
\label{eq:qc-pure-relations}
\end{equation}
Indeed, there exists a basis where $\rho$ is diagonal, with one eigenvalue equal to one and all other eigenvalues zero. In this basis the root $\tilde\Psi = \rho^{1/2}$ is easily taken, leading to the relations \eqref{eq:qc-pure-relations}. These relations are preserved by a change of basis and therefore hold in an arbitrary basis.

For a pure state density matrix $\rho(\varphi,\varphi')$ there exists a complex quantum wave function $\psi(\varphi)$,
\begin{equation}
\rho(\varphi,\varphi') = \psi(\varphi)\, \psi^*(\varphi') \,, \qquad \int \tilde D\varphi\; |\psi(\varphi)|^2 = 1 \,.
\label{eq:qc-rho-pure}
\end{equation}
In turn, the wave function $\tilde\Psi(\varphi,\chi)$ takes a direct product form,
\begin{equation}
\tilde\Psi(\varphi,\chi) = \psi(\varphi)\, \psi^*(\chi) \,.
\label{eq:qc-psi-product}
\end{equation}

For our discussion of equilibrium states only the quantum ground state is a pure quantum state. In the limit $\beta\to\infty$, $T\to0$, only one term contributes in eq.~\eqref{eq:psi-thermal-mode}, \eqref{eq:Zp-thermal} to the sum over $m$. For the classical thermal equilibrium of the harmonic oscillator one finds from eq.~\eqref{eq:psicl-zzeta} the quantum density matrix
\begin{equation}
\rho(x,x') = \sqrt{\frac{8}{\pi\beta g}}\, \exp\left\{-\left[\frac{\beta\omega^2 g}{32}(x-x')^2 + \frac{2}{\beta g}(x+x')^2\right]\right\} \,, \qquad g = 1 + \frac{16}{\omega^2\beta^2} \,.
\label{eq:qc-rho-oscillator}
\end{equation}
For the free classical field theory $\rho$ becomes a product of density matrices for the momentum modes, with the replacement $x\to\varphi_j(\vec p)$, $x'\to\varphi_j'(\vec p)$, and $\omega^2 = m^2+\vec p^{\,2}$. Towards the classical ground state for $\beta\to\infty$ the off-diagonal elements of the density matrix vanish rapidly for $(x-x')^2 \gg 32/(\beta\omega^2)$. In contrast, the diagonal elements for $x'=x$ are almost constant for a large range $x^2 \ll \beta g/8$,
\begin{equation}
\rho(x,x) = \sqrt{\frac{8}{\pi\beta g}}\, \exp\left(-\frac{8 x^2}{\beta g}\right) \,.
\label{eq:qc-rho-diagonal-oscillator}
\end{equation}
For $\beta\to\infty$ this is far from a pure quantum state. In eq.~\eqref{eq:qc-rho-diagonal-oscillator} the dependence on $\omega$ arises only from $g$.

\subsection{The classical-quantum map}
\label{sec:cq-map}

We have described a general ``classical-quantum map'' from a classical probability distribution to a quantum density matrix. More precisely, it is a map from the classical wave function $q$ to the quantum density matrix, since the sign of $q$ must be fixed. The map backwards from $\tilde\Psi$ to $q$ is a Fourier transform of the ``relative field'' $\varphi-\chi$ to $\pi$, keeping the ``average field'' $\frac12(\varphi+\chi) = \sigma$ fixed. Applying the same Fourier transform to the density matrix $\rho$ amounts to the Wigner function $W[\sigma,\pi]$. The Wigner function is, in general, not positive definite, in contrast to the probability distribution $w=q^2$. It remains positive, however, for the quantum thermal equilibrium state.

\begin{figure}[ht]
\centering
\includegraphics[width=0.85\textwidth]{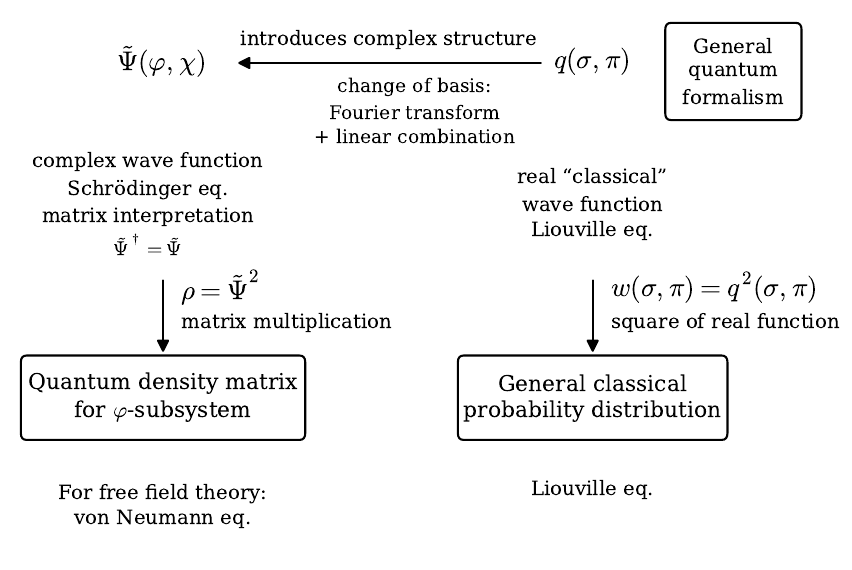}
\caption{Maps between $q$, $\tilde\Psi$, $w$, $\rho$. We also indicate the form of the evolution equation.}
\label{fig:maps}
\end{figure}

This type of classical-quantum map is defined for an arbitrary classical wave function $q[\sigma,\pi]$, independent of the dynamics and independent of the exact form of the classical field equation. We depict the various maps of this construction in fig.~\ref{fig:maps}.

The practical use of this quantum density matrix is often limited mainly to free theories. The reason is that, in general, $\rho$ does not evolve according to the von Neumann equation. Only free theories, or more generally, theories with a separable Hamiltonian $H_L$, realize the unitary evolution of $\rho$ according to the von Neumann equation.

The time evolution of the wave function $\tilde\Psi(\varphi,\chi)$ obeys the generalized Schr\"odinger equation
\begin{equation}
i\partial_t \tilde\Psi(\varphi,\chi) = \int \tilde D\varphi'\, \tilde D\chi'\; H_L(\varphi,\chi;\varphi',\chi')\, \tilde\Psi(\varphi',\chi') \,.
\label{eq:cq-schroedinger-generalized}
\end{equation}
This translates for the density matrix to
\begin{equation}
\begin{aligned}
i\partial_t \rho(\varphi,\varphi') = \int \tilde D\chi\, \tilde D\varphi''\, \tilde D\chi''\, \Big[ &H_L(\varphi,\chi;\varphi'',\chi'')\, \tilde\Psi(\varphi'',\chi'')\, \tilde\Psi(\chi,\varphi') \\
+\, &\tilde\Psi(\varphi,\chi)\, H_L(\chi,\varphi';\chi'',\varphi'')\, \tilde\Psi(\chi'',\varphi'') \Big] \,.
\end{aligned}
\label{eq:cq-rho-evolution}
\end{equation}
For classical field equations
\begin{equation}
\partial_t \sigma = \pi \,, \qquad \partial_t \pi = F(\sigma) \,,
\label{eq:cq-classical-eom}
\end{equation}
one has
\begin{equation}
H_L = -\int \left\{ \frac12 \left( \frac{\partial^2}{\partial\varphi^2} - \frac{\partial^2}{\partial\chi^2} \right) + W(\varphi,\chi) \right\} ,
\label{eq:cq-HL-W}
\end{equation}
with
\begin{equation}
W(\varphi,\chi) = (\varphi-\chi)\, F\!\left(\frac{\varphi+\chi}{2}\right) = -W(\chi,\varphi) \,.
\label{eq:cq-W-def}
\end{equation}
Here we omit indices or arguments of functions, and $\int$ denotes appropriate index sums or integrals. Eq.~\eqref{eq:cq-rho-evolution} becomes
\begin{equation}
i\partial_t \rho(\varphi,\varphi') = -\frac12 \left( \frac{\partial^2}{\partial\varphi^2} - \frac{\partial^2}{\partial\varphi'^2} \right) \rho(\varphi,\varphi') - \int \tilde D\chi \left( W(\varphi,\chi) - W(\varphi',\chi) \right) \tilde\Psi(\varphi,\chi)\, \tilde\Psi(\chi,\varphi') \,.
\label{eq:cq-rho-W}
\end{equation}
In general, this expression is not linear in $\rho$. For general $W$ the second term in eq.~\eqref{eq:cq-rho-W} involves $\tilde\Psi = \rho^{1/2}$.

For a free theory $F$ is linear in $\sigma$, resulting in
\begin{equation}
W(\varphi,\chi) - W(\varphi',\chi) = W(\varphi,\varphi') \,,
\label{eq:cq-W-free}
\end{equation}
such that the r.h.s.\ of eq.~\eqref{eq:cq-rho-W} is linear in $\rho$. In this case the Hamiltonian can be decomposed,
\begin{equation}
H_L(\varphi,\chi;\varphi'',\chi'') = H^{(q)}(\varphi,\varphi'')\, \delta(\chi-\chi'') - H^{(q)}(\chi'',\chi)\, \delta(\varphi-\varphi'') \,,
\label{eq:cq-HL-factorizing}
\end{equation}
and one finds
\begin{equation}
i\partial_t \tilde\Psi = [H^{(q)}, \tilde\Psi] \,.
\label{eq:cq-psi-commutator}
\end{equation}
This entails the von Neumann equation for the density matrix
\begin{equation}
i\partial_t \rho = [H^{(q)}, \rho] \,.
\label{eq:cq-vonNeumann}
\end{equation}
This setting is realized for general free classical field theories.

For classical field theories with interactions the decomposition \eqref{eq:cq-HL-factorizing} is not realized. The quantum Hamiltonian $H^{(q)}$ is no longer conserved, only $H_L$ is a conserved quantity. For a realistic quantum system one would like a generalization of a conserved $H^{(q)}(\varphi)$, either by integrating out the mirror field $\chi$ \cite{CWQF}, or by some particular form of interaction. Furthermore, there often exist other classical-quantum maps. They typically map to a closed subsystem, associated to a restricted subset of classical probability distributions. The quantum hydrogen atom as a subsystem of the probabilistic particle in the Coulomb potential is an example \cite{CWQP}. For a closed subsystem $H^{(q)}$ becomes the projection of $H_L$ on the subsystem.

We emphasize that the general quantum formalism based on the wave functions $q$ or $\tilde\Psi$ is valid for general classical field equations \eqref{eq:cq-classical-eom}, including interactions. Only the particular map to a subsystem for which $\rho$ obeys the von Neumann equation is restricted to free theories.

\subsection{Comparison with quantum optics}
\label{sec:quantum-optics}

In quantum optics it is well known that the Planck spectrum can be obtained from a positive probability distribution. This holds for the Wigner function \cite{WIG}, the Husimi function \cite{HUS} and the Glauber-Sudarshan $P$-function \cite{GLA,SUD}. The Planck spectrum has also been derived within stochastic electrodynamics from classical electromagnetic fields with classical zero-point radiation \cite{MAR1,BOY1,BOY2,BOY3,QDICE}. The new point of the present note is not the existence of a positive probability distribution for the description of the Planck spectrum, but rather its realization as a probability distribution for classical fields. It is therefore appropriate to compare our setting based on $w[\sigma,\pi]$ with the probability distributions used in quantum optics.

The Wigner function, the Husimi function and the $P$-function are all derived from a quantum field theory for free electromagnetic fields. The probability distributions of relevance here are products of probability distributions for the different $\vec p$-modes. It is therefore sufficient to discuss the issue for a single harmonic oscillator corresponding to a given $\vec p$-mode. As in sect.~\ref{sec:harmonic} it is characterized by a frequency $\omega$ which will depend on $\vec p$. Our description of the probabilistic classical harmonic oscillator contains the quantum harmonic oscillator as a subsystem. We can therefore perform all steps of the quantum theory. This includes the partial Fourier transform of the quantum density matrix to the Wigner function. We can also perform a change of basis to the coherent state basis and the corresponding map to the density matrix in this basis. From there one can compute the $P$-function or the Husimi function. The positive probabilities used in quantum optics are contained in our approach.

The Wigner function and the $P$-function do not remain positive for a general quantum density matrix, in contrast to the classical probability distribution $w(z,v)$. The Husimi function is a positive function and in that sense perhaps closest to our setting. In the general case the extraction of expectation values of observables from the Husimi function is rather cumbersome due to the required anti-normal ordering of operators. The Husimi function consists of the diagonal elements of the density matrix in the coherent state basis. The off-diagonal elements have to be reconstructed by analyticity requirements. In contrast, in our approach the computation of all expectation values of both classical and statistical observables is much more direct. In terms of the real wave function $q(z,v)$, or the associated $\tilde\psi(x,y)$, the expectation values obtain from the usual quantum rule by associating operators to observables. Key statistical observables are represented by simple derivative-operators. Once one has made the step to describe the time-local probabilistic information by a wave function which is the square root of the probability distribution, everything follows simple well established procedures. In particular, one can define probabilities from the expectation values, both for classical and statistical observables. For these probabilities the Born rule follows for both types of observables \cite{CWDS}.

Statistical observables are represented by operators which do not commute with the ones for classical observables. As a consequence, simultaneous probabilities for the possible measurement values of a pair of classical and statistical observable do not exist. The existence of simultaneous probabilities is a central assumption for the derivation of Bell's inequalities. One can, nevertheless, define correlations for measurements of such observables, based on conditional probabilities. Those do not need to obey Bell's inequalities \cite{CWPW}. This often generalizes to pairs of statistical observables. If the operators associated to a pair of observables do not commute, Bell's inequalities do not have to hold for correlations of these observables.

For the purpose of this paper of comparing the classical and quantum thermal equilibrium the probability distribution for classical fields (of phase space variables $z,v$) for both equilibrium states is crucial. The Husimi function can be used for the quantum equilibrium state. The notion of a different classical equilibrium state is not defined for this function.

\subsection{Time dependence of correlation functions}
\label{sec:time-dependence-correlations}

The classical thermal equilibrium and the quantum thermal equilibrium are both candidates for a state which the classical probabilistic field theory may reach in the evolution for a long time. Actually, it is not necessary that the exact thermal equilibrium probability distributions are reached. It is sufficient that the low correlation functions approximate those computed from the equilibrium states. Additional information may be stored in correlations of very high order that are inaccessible in practice. One would like to know which one of the two equilibrium states is reached in the asymptotic time limit if small interactions are added. The result may depend on the type of interactions. This question matters only for initial states with a sufficiently large value of $\langle E^{(cl)} \rangle$ such that both equilibrium states are possible. Since $\langle E^{(cl)} \rangle$ is a conserved quantity, an initial value below the one of the quantum ground state forbids an asymptotic quantum equilibrium state.

The equal-time correlation functions can be extracted from the time-dependent effective action $\Gamma[\bar\sigma,t]$, which is the generating functional for the one-particle irreducible equal-time correlation functions \cite{CWET1}. For our classical field theory the variables $\bar\sigma_m$ are given by $(\sigma(\vec x),\pi(\vec x))$. One considers a formal classical field equation for $\tilde\sigma$,
\begin{equation}
\partial_t \tilde\sigma_m = F_m[\tilde\sigma] \,,
\label{eq:ET-field-eq}
\end{equation}
given by eq.~\eqref{eq:field-eom} in our case, and extended in case of interactions or inhomogeneities. The equal-time effective action obeys an exact evolution equation \cite{CWET1}
\begin{equation}
\partial_t \Gamma[\bar\sigma] = -\frac{\partial\Gamma[\bar\sigma]}{\partial\bar\sigma_m}\, \hat F_m[\bar\sigma] \,.
\label{eq:ET-Gamma-evolution}
\end{equation}
Here $\bar\sigma = \langle\tilde\sigma\rangle$, and $\hat F_m$ can be computed from $F_m$. For suitable truncations of $\Gamma$ they can be solved numerically and compared with direct simulations of classical field equations with probabilistic initial conditions \cite{BOWE,ABW,TSE,OZB}.

For linear field equations the functionals $F_m$ and $\hat F_m$ are identical,
\begin{equation}
F_m[\tilde\sigma] = f_{mn}\, \tilde\sigma_n \,, \qquad \hat F_m[\bar\sigma] = f_{mn}\, \bar\sigma_n \,.
\label{eq:ET-linear}
\end{equation}
Expanding $\Gamma$ in powers of $\bar\sigma$, the evolution equation \eqref{eq:ET-Gamma-evolution} does not mix different powers. The 1PI correlation functions of a given order evolve independently. In particular, the second functional derivative $\Gamma^{(2)}$ is the inverse two-point function. Interactions will mix the correlation functions of different order.

The equal-time effective action can also be formulated for a quantum field theory \cite{CWET2,NAC}. In our case $\Gamma[\bar\varphi,t]$, $\bar\varphi = (\varphi(\vec x),\eta(\vec x))$, describes the correlation functions for $\varphi$ and $\eta$. For a free quantum field theory the evolution equation for the quantum field theory is the same as for the classical theory. In the presence of interactions the generator $\hat F_m[\bar\varphi]$ receives additional terms which arise from the non-vanishing commutator $[\hat\varphi,\hat\eta]$.

One concludes in our case that the evolution equation for the correlation $\langle\sigma(\vec x)\sigma(\vec y)\rangle$ is the same as for $\langle\varphi(\vec x)\varphi(\vec y)\rangle$. They differ only by different initial conditions. For our two equilibrium states both two-point functions are time-independent. Linear field equations extend to the presence of inhomogeneities, as fields in a potential or coupling to external ``background fields''. Also the initial conditions for the correlation functions can be inhomogeneous. Nevertheless, a decision on which asymptotic state is preferred cannot be done on the linear level. Equilibration needs interactions.

An important role is played by conserved quantities, or conserved correlation functions. A prime example is the classical energy, which remains conserved in the presence of interactions. Arbitrary correlations of the classical energy, as the energy dispersion $\Delta E^2$, are conserved. One may start with the correlation functions of the quantum ground state and switch on an interaction with a small strength $\lambda$. The relative energy dispersion $\Delta E^2/\bar E^2$ will not change during the evolution. For sufficiently small $\lambda$ the relative energy dispersion of the classical thermal equilibrium is given by $1+a\lambda$. It cannot be reached by an initial $\Delta E^2/\bar E^2$ close to $16/15$ as appropriate for the quantum ground state. This is an example for which the precise Boltzmann distribution for the classical thermal equilibrium cannot be reached for asymptotic times. The system could still reach a type of classical thermal equilibrium with modified energy dispersion \cite{ABW}. For large $N$ this may be almost indistinguishable from the Boltzmann distribution.

In the presence of symmetries there exist other conserved quantities based on classical or statistical observables. They may include the classical and the quantum angular momentum, or conserved charges in case of internal symmetries. (Note that conserved charges are typically statistical observables.) Conserved charges can obstruct the approach to the classical thermal equilibrium. This does not imply that some quantum thermal equilibrium is reached. Still, if the quantum thermal equilibrium is compatible with the conserved correlations it is a valid candidate for the asymptotic state.

\section{Discussion}
\label{sec:discussion}

The probabilistic description of a free classical field theory contains a quantum field theory as a subsystem. The quantum and the classical thermal equilibrium correspond to two different classical probability distributions. There is no fundamental conceptual difference between a probability distribution describing the Planck spectrum and a probability distribution exhibiting the Rayleigh-Jeans divergence of the Boltzmann distribution for classical fields. This extends to the Maxwell equations in vacuum. One can find the Planck spectrum of the black body radiation by assuming an equilibrium distribution different from the Boltzmann distribution. The classical probability distribution for the quantum thermal equilibrium describes a highly fluctuating state even in the zero temperature limit. This differs from the Boltzmann distribution for which the zero temperature limit corresponds to sharply vanishing fields.

From the point of view of this classical probabilistic setting an important question arises: Can one explain why observation sees the Planck spectrum and not the Rayleigh-Jeans spectrum? On the level of a free field theory both equilibrium states are given by static and homogeneous probability distributions. As an important difference the Planck spectrum remains finite in the limit of infinite resolution or infinite momentum cutoff. Only the unobservable ground state energy diverges. This contrasts with the Rayleigh-Jeans spectrum.

For the quantum equilibrium the situation is similar to correlation functions in quantum field theory: there exist particular correlation functions (typically for renormalized fields) which remain finite in the limit of infinite momentum cutoff, whereas other correlation functions diverge. Only the UV-finite correlation functions can be associated to observables. This point favors the Planck spectrum.

One would like to find out if in the presence of interactions an initial state with sufficient fluctuations approaches the quantum thermal equilibrium for asymptotic times $t\to\infty$. This issue may depend on the type of interaction. One may consider an interaction characterized by a small parameter $\lambda$. In the limit of very small $\lambda$ the correlation functions computed for the equilibrium states of the free classical field theory are expected to be a good approximation for possible asymptotic states of the interacting theory. One could then judge from the time evolution of correlation functions in the interacting theory if one finds a tendency to approach the classical or quantum thermal equilibrium. In practice one may start initially from the probability distribution for the quantum ground state or quantum thermal equilibrium of the free theory. Switching on a small interaction one can follow the time evolution of the correlation functions and judge if for small momenta the enhancement characteristic for the classical thermal equilibrium sets in, or if the correlation functions remain close to the ones for the quantum ground state.

\begin{figure}[tb]
\centering
\includegraphics[width=0.62\textwidth]{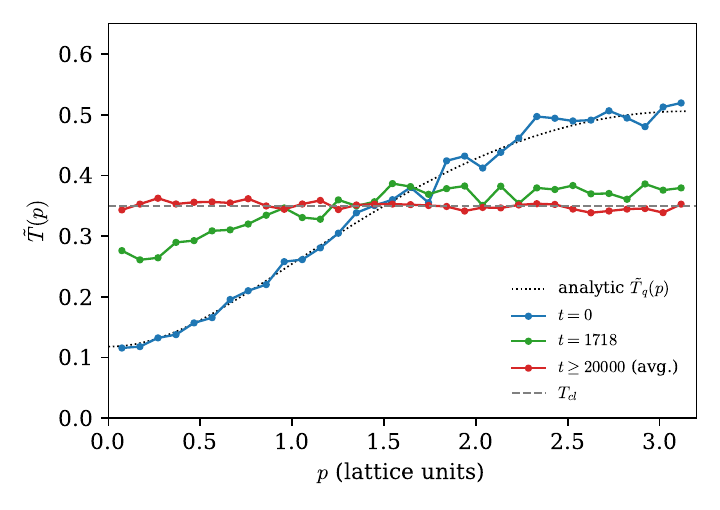}
\caption{Time evolution of the mode temperature $\tilde T(p)$ in a classical lattice simulation of a complex scalar field $\Phi=(\phi_1+i\phi_2)/\sqrt2$ with U(1) symmetry in one space dimension. The interaction is $\frac{\lambda}{6}(\Phi^*\Phi)^2=\frac{\lambda}{24}(\phi_a\phi_a)^2$ with $\lambda=1$; the lattice has $N=128$ sites with lattice spacing $a=1$. The ensemble consists of $M=256$ realizations of the complex field. Initially it is drawn from the quantum thermal equilibrium of the free theory with $T_q=0.1$ and effective mass $m=0.3$ (the bare mass is chosen such that the Hartree-corrected mass equals $m$ at $t=0$). Shown are $\tilde T(p)$ at $t=0$ (blue, the dotted line is the analytic $\tilde T_q(p)$), at $t\approx 1700$ (green), and averaged over $t\geq 20000$ (red); the dashed line marks the flat classical value $T_{cl}$. Momenta $p$ are in lattice units. A real scalar field with $\phi^4$ interaction gives similar results.}
\label{fig:lattice}
\end{figure}

As an example, a small numerical simulation of the $1+1$-dimensional classical field theory for a complex scalar field $\Phi$ with quartic interaction $\frac{\lambda}{6}(\Phi^*\Phi)^2$ shows an approach towards the classical thermal equilibrium. This can be seen in fig.~\ref{fig:lattice} where the initial distribution of the free quantum thermal equilibrium approaches for large $t$ the flat distribution of the classical thermal equilibrium, in accordance with similar results in the past \cite{BMS95,ABW,BOWE,AB2002,BBW04,MT04}. This simple model demonstrates how a selection by dynamics can proceed. The question arises what happens in the limit $\Lambda\to\infty$ for which the Rayleigh-Jeans divergence no longer allows for a regular behavior. For $\Lambda\to\infty$ the transition frequency $\bar\omega\approx\Lambda/2$ (in one space dimension for $\omega=|\vec p|$; for the lattice dispersion $\omega=(2/a)|\sin(pa/2)|$ one finds $\bar\omega\approx 4/(\pi a)$) where the quantum spectrum crosses the classical spectrum moves to infinity. For the low momentum modes with $\omega(\vec p)\ll\bar\omega$ the energy needed in order to ``heat'' $\tilde T(\vec p)$ towards $\tilde T_{cl}$ has to come from higher and higher momenta. For an increasing cutoff at fixed quartic coupling, one observes that the time for the deviation from the quantum thermal equilibrium, measured in units of the inverse cutoff, increases substantially. This poses the question if this time could diverge in physical units for cutoff going to infinity, such that the system remains close to the quantum thermal equilibrium. An investigation of this question requires to take a proper continuum limit.

\begin{figure}[tb]
\centering
\includegraphics[width=0.62\textwidth]{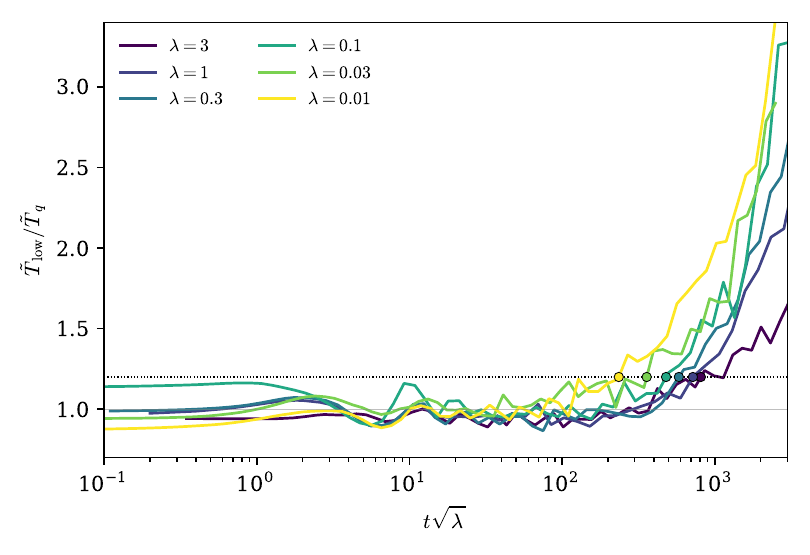}
\caption{Dependence on the coupling in a classical lattice simulation of a real scalar field in one space dimension with interaction $\frac{\lambda}{24}\phi^4$. All dimensionful quantities are scaled with $\sqrt\lambda$: the effective mass is $m=0.3\sqrt\lambda$, the quantum temperature is $T_q=0.1\sqrt\lambda$, and the box length is $L=128/\sqrt\lambda$, such that the momenta $p/\sqrt\lambda=2\pi n/128$ are the same for all $\lambda$. The lattice spacing is fixed, $a=1$, so that the ultraviolet cutoff $\Lambda=\pi/a$ corresponds to $\Lambda/\sqrt\lambda=1.8,\,3.1,\,5.7,\,9.9,\,18,\,31$ for $\lambda=3,\,1,\,0.3,\,0.1,\,0.03,\,0.01$ (with $N=74,\,128,\,234,\,405,\,739,\,1280$ lattice sites). The ensemble consists of $M=96$ realizations drawn initially from the quantum thermal equilibrium of the free theory; the bare mass is chosen such that the Hartree-corrected mass equals $m$ at $t=0$. Shown is $\tilde T_{\rm low}$, the mode temperature $\tilde T(p)$ averaged over the four lowest nonzero momenta ($p/\sqrt\lambda\leq 0.2$), in units of the quantum value $\tilde T_q=\frac{\omega}{4}\coth\frac{\omega}{4T_q}$ averaged over the same modes, as a function of $t\sqrt\lambda$. The dots mark the time $t_{20}$ at which $\tilde T$ deviates from $\tilde T_q$ by more than $20\%$ (dotted line), with $t_{20}\sqrt\lambda\approx 810,\,720,\,580,\,480,\,360,\,235$ for decreasing $\lambda$. The statistical uncertainty of each curve is about $10\%$. The runs for $\lambda=0.03$ and $0.01$ end at $t\sqrt\lambda=2500$.}
\label{fig:scaled}
\end{figure}

In two space-time dimensions the field $\varphi$ is dimensionless, such that $\lambda$ and $m^2$ have dimension mass$^2$. For the continuum limit one may use $\sqrt\lambda$ in order to fix the physical scale. The continuum limit corresponds then to $\lambda/\Lambda^2\to0$. The renormalized mass term $m_R^2$ should be fixed in units of $\lambda$, i.e.\ $m_R^2/\lambda$ fixed. (Here $m_R^2$ is related to the bare mass term $m_0^2$ by a ``quantum correction'' $\sim\lambda\ln(\Lambda^2/m_R^2)$.) The temperature $T_q$ should be taken in units of $\sqrt\lambda$, e.g.\ $T_q/\sqrt\lambda$ fixed. The box length $L$ should be fixed in units of $1/\sqrt\lambda$, i.e.\ $L\sqrt\lambda$ fixed. Also physical time should be measured in units of $1/\sqrt\lambda$. With $a$ the lattice distance this means $\sqrt\lambda\,t=(a\sqrt\lambda)\,t/a$. For a fixed physical time interval the lattice time interval $\Delta t/a$ increases $\sim(a\sqrt\lambda)^{-1}\sim\Lambda/\sqrt\lambda$. Finally, constant physical momenta $p/\sqrt\lambda$ correspond to $pa\sim a\sqrt\lambda\sim\sqrt\lambda/\Lambda$. Instead of varying $\lambda$ we can take a fixed lattice distance $a$ and vary the physical quantities as described. We show in fig.~\ref{fig:scaled} for a real scalar field the evolution of $\tilde T(p)/\tilde T_q(p)$ for the smallest momenta, with the Hartree-corrected effective mass $m=0.3\sqrt\lambda$ representing $m_R$ up to finite corrections. The physical time that it takes to deviate substantially (by more than $20\%$) from the quantum equilibrium somewhat decreases with increasing $\Lambda/\sqrt\lambda$. For the chosen parameters, the physical time for the deviation from the quantum thermal equilibrium is finite. We observe, however, that for increasing $m/\sqrt\lambda$ the physical time for the deviation from the quantum thermal equilibrium increases rapidly (at $T_q/\sqrt\lambda=0.1$ and $\Lambda/\sqrt\lambda=\pi$ one finds $t_{20}\sqrt\lambda\approx 700$ for $m/\sqrt\lambda=0.3$ and $\approx 9000$ for $0.45$, while no deviation is seen up to $t\sqrt\lambda=3\times10^4$ for $0.6$). This poses the question if there are parameters for which this time becomes very large.

Our $1+1$ dimensional simulation should only be seen as a demonstration of possible ways to answer the dynamical question. It may well be that the answer comes out differently in other dimensions and for different interactions. For electromagnetic fields the issue may depend on the properties of particles to which the field couples. A classical probabilistic setting for electrons needs a description of fermions by a classical probability distribution. This could proceed by the equivalence of fermionic quantum field theories with generalized Ising models \cite{CWFI,CWCPT}. If electrons behave effectively as quantum electrons, an approach to the quantum thermal equilibrium seems likely. It is a challenging question if the observed Planck spectrum of the black body radiation is finally predicted by a classical probabilistic field theory. If this possibility is realized, this may profoundly change the view of quantum mechanics. Quantum mechanics may emerge from a classical probabilistic setting not only on the conceptual level \cite{CWPW,CWQP1,CWQF}, but also as a result of dynamics.

\begin{acknowledgments}
The author would like to thank J.~Berges for stimulating discussions. He has worked with Claude for typesetting, checks, figures, simulation, reference search, and critical assessment.
\end{acknowledgments}

\nocite{*}
\bibliography{refs}

\begin{thebibliography}{71}%
\makeatletter
\providecommand \@ifxundefined [1]{%
 \@ifx{#1\undefined}
}%
\providecommand \@ifnum [1]{%
 \ifnum #1\expandafter \@firstoftwo
 \else \expandafter \@secondoftwo
 \fi
}%
\providecommand \@ifx [1]{%
 \ifx #1\expandafter \@firstoftwo
 \else \expandafter \@secondoftwo
 \fi
}%
\providecommand \natexlab [1]{#1}%
\providecommand \enquote  [1]{``#1''}%
\providecommand \bibnamefont  [1]{#1}%
\providecommand \bibfnamefont [1]{#1}%
\providecommand \citenamefont [1]{#1}%
\providecommand \href@noop [0]{\@secondoftwo}%
\providecommand \href [0]{\begingroup \@sanitize@url \@href}%
\providecommand \@href[1]{\@@startlink{#1}\@@href}%
\providecommand \@@href[1]{\endgroup#1\@@endlink}%
\providecommand \@sanitize@url [0]{\catcode `\\12\catcode `\$12\catcode
  `\&12\catcode `\#12\catcode `\^12\catcode `\_12\catcode `\%12\relax}%
\providecommand \@@startlink[1]{}%
\providecommand \@@endlink[0]{}%
\providecommand \url  [0]{\begingroup\@sanitize@url \@url }%
\providecommand \@url [1]{\endgroup\@href {#1}{\urlprefix }}%
\providecommand \urlprefix  [0]{URL }%
\providecommand \Eprint [0]{\href }%
\providecommand \doibase [0]{https://doi.org/}%
\providecommand \selectlanguage [0]{\@gobble}%
\providecommand \bibinfo  [0]{\@secondoftwo}%
\providecommand \bibfield  [0]{\@secondoftwo}%
\providecommand \translation [1]{[#1]}%
\providecommand \BibitemOpen [0]{}%
\providecommand \bibitemStop [0]{}%
\providecommand \bibitemNoStop [0]{.\EOS\space}%
\providecommand \EOS [0]{\spacefactor3000\relax}%
\providecommand \BibitemShut  [1]{\csname bibitem#1\endcsname}%
\let\auto@bib@innerbib\@empty
%</preamble>
\bibitem [{\citenamefont {Mather}\ \emph {et~al.}(1994)\citenamefont {Mather}
  \emph {et~al.}}]{CMB1}%
  \BibitemOpen
  \bibfield  {author} {\bibinfo {author} {\bibfnamefont {J.~C.}\ \bibnamefont
  {Mather}} \emph {et~al.},\ }\bibfield  {title} {\bibinfo {title} {Measurement
  of the cosmic microwave background spectrum by the {COBE} {FIRAS}
  instrument},\ }\href {https://doi.org/10.1086/173574} {\bibfield  {journal}
  {\bibinfo  {journal} {Astrophys. J.}\ }\textbf {\bibinfo {volume} {420}},\
  \bibinfo {pages} {439} (\bibinfo {year} {1994})}\BibitemShut {NoStop}%
\bibitem [{\citenamefont {{Planck Collaboration}}\ \emph
  {et~al.}(2020)\citenamefont {{Planck Collaboration}}, \citenamefont {Aghanim}
  \emph {et~al.}}]{CMB2}%
  \BibitemOpen
  \bibfield  {author} {\bibinfo {author} {\bibnamefont {{Planck
  Collaboration}}}, \bibinfo {author} {\bibfnamefont {N.}~\bibnamefont
  {Aghanim}}, \emph {et~al.},\ }\bibfield  {title} {\bibinfo {title} {{P}lanck
  2018 results. {V}. {CMB} power spectra and likelihoods},\ }\href
  {https://doi.org/10.1051/0004-6361/201936386} {\bibfield  {journal} {\bibinfo
   {journal} {Astron. Astrophys.}\ }\textbf {\bibinfo {volume} {641}},\
  \bibinfo {pages} {A5} (\bibinfo {year} {2020})},\ \Eprint
  {https://arxiv.org/abs/1907.12875} {arXiv:1907.12875 [astro-ph.CO]}
  \BibitemShut {NoStop}%
\bibitem [{\citenamefont {Le~Bellac}(1996)}]{TFT1}%
  \BibitemOpen
  \bibfield  {author} {\bibinfo {author} {\bibfnamefont {M.}~\bibnamefont
  {Le~Bellac}},\ }\href {https://doi.org/10.1017/CBO9780511721700} {\emph
  {\bibinfo {title} {Thermal Field Theory}}}\ (\bibinfo  {publisher} {Cambridge
  University Press},\ \bibinfo {address} {Cambridge},\ \bibinfo {year}
  {1996})\BibitemShut {NoStop}%
\bibitem [{\citenamefont {Kraemmer}\ and\ \citenamefont {Rebhan}(2004)}]{TFT2}%
  \BibitemOpen
  \bibfield  {author} {\bibinfo {author} {\bibfnamefont {U.}~\bibnamefont
  {Kraemmer}}\ and\ \bibinfo {author} {\bibfnamefont {A.}~\bibnamefont
  {Rebhan}},\ }\bibfield  {title} {\bibinfo {title} {Advances in perturbative
  thermal field theory},\ }\href {https://doi.org/10.1088/0034-4885/67/3/R05}
  {\bibfield  {journal} {\bibinfo  {journal} {Rept. Prog. Phys.}\ }\textbf
  {\bibinfo {volume} {67}},\ \bibinfo {pages} {351} (\bibinfo {year} {2004})},\
  \Eprint {https://arxiv.org/abs/hep-ph/0310337} {arXiv:hep-ph/0310337
  [hep-ph]} \BibitemShut {NoStop}%
\bibitem [{\citenamefont {Laine}\ and\ \citenamefont {Vuorinen}(2016)}]{TFT3}%
  \BibitemOpen
  \bibfield  {author} {\bibinfo {author} {\bibfnamefont {M.}~\bibnamefont
  {Laine}}\ and\ \bibinfo {author} {\bibfnamefont {A.}~\bibnamefont
  {Vuorinen}},\ }\href {https://doi.org/10.1007/978-3-319-31933-9} {\emph
  {\bibinfo {title} {Basics of Thermal Field Theory}}},\ \bibinfo {series}
  {Lecture Notes in Physics}, Vol.\ \bibinfo {volume} {925}\ (\bibinfo
  {publisher} {Springer International Publishing},\ \bibinfo {address} {Cham},\
  \bibinfo {year} {2016})\ \Eprint {https://arxiv.org/abs/1701.01554}
  {arXiv:1701.01554 [hep-ph]} \BibitemShut {NoStop}%
\bibitem [{\citenamefont {Polarski}\ and\ \citenamefont
  {Starobinsky}(1996)}]{PSTA}%
  \BibitemOpen
  \bibfield  {author} {\bibinfo {author} {\bibfnamefont {D.}~\bibnamefont
  {Polarski}}\ and\ \bibinfo {author} {\bibfnamefont {A.~A.}\ \bibnamefont
  {Starobinsky}},\ }\bibfield  {title} {\bibinfo {title} {Semiclassicality and
  decoherence of cosmological perturbations},\ }\href
  {https://doi.org/10.1088/0264-9381/13/3/006} {\bibfield  {journal} {\bibinfo
  {journal} {Class. Quantum Grav.}\ }\textbf {\bibinfo {volume} {13}},\
  \bibinfo {pages} {377} (\bibinfo {year} {1996})},\ \Eprint
  {https://arxiv.org/abs/gr-qc/9504030} {arXiv:gr-qc/9504030} \BibitemShut
  {NoStop}%
\bibitem [{\citenamefont {Son}(1996)}]{DTS}%
  \BibitemOpen
  \bibfield  {author} {\bibinfo {author} {\bibfnamefont {D.~T.}\ \bibnamefont
  {Son}},\ }\href@noop {} {\bibinfo {title} {Classical preheating and
  decoherence}} (\bibinfo {year} {1996}),\ \Eprint
  {https://arxiv.org/abs/hep-ph/9601377} {arXiv:hep-ph/9601377 [hep-ph]}
  \BibitemShut {NoStop}%
\bibitem [{\citenamefont {Khlebnikov}\ and\ \citenamefont
  {Tkachev}(1996)}]{YK}%
  \BibitemOpen
  \bibfield  {author} {\bibinfo {author} {\bibfnamefont {S.~Y.}\ \bibnamefont
  {Khlebnikov}}\ and\ \bibinfo {author} {\bibfnamefont {I.~I.}\ \bibnamefont
  {Tkachev}},\ }\bibfield  {title} {\bibinfo {title} {Classical decay of the
  inflaton},\ }\href {https://doi.org/10.1103/PhysRevLett.77.219} {\bibfield
  {journal} {\bibinfo  {journal} {Phys. Rev. Lett.}\ }\textbf {\bibinfo
  {volume} {77}},\ \bibinfo {pages} {219} (\bibinfo {year} {1996})},\ \Eprint
  {https://arxiv.org/abs/hep-ph/9603378} {arXiv:hep-ph/9603378} \BibitemShut
  {NoStop}%
\bibitem [{\citenamefont {Gelis}\ \emph {et~al.}(2007)\citenamefont {Gelis},
  \citenamefont {Lappi},\ and\ \citenamefont {Venugopalan}}]{GLV}%
  \BibitemOpen
  \bibfield  {author} {\bibinfo {author} {\bibfnamefont {F.}~\bibnamefont
  {Gelis}}, \bibinfo {author} {\bibfnamefont {T.}~\bibnamefont {Lappi}},\ and\
  \bibinfo {author} {\bibfnamefont {R.}~\bibnamefont {Venugopalan}},\
  }\bibfield  {title} {\bibinfo {title} {High energy scattering in quantum
  chromodynamics},\ }\href {https://doi.org/10.1142/S0218301307008331}
  {\bibfield  {journal} {\bibinfo  {journal} {Int. J. Mod. Phys. E}\ }\textbf
  {\bibinfo {volume} {16}},\ \bibinfo {pages} {2595} (\bibinfo {year}
  {2007})},\ \Eprint {https://arxiv.org/abs/0708.0047} {arXiv:0708.0047
  [hep-ph]} \BibitemShut {NoStop}%
\bibitem [{\citenamefont {Fukushima}\ \emph {et~al.}(2007)\citenamefont
  {Fukushima}, \citenamefont {Gelis},\ and\ \citenamefont {McLerran}}]{FGL}%
  \BibitemOpen
  \bibfield  {author} {\bibinfo {author} {\bibfnamefont {K.}~\bibnamefont
  {Fukushima}}, \bibinfo {author} {\bibfnamefont {F.}~\bibnamefont {Gelis}},\
  and\ \bibinfo {author} {\bibfnamefont {L.}~\bibnamefont {McLerran}},\
  }\bibfield  {title} {\bibinfo {title} {Initial singularity of the little
  bang},\ }\href {https://doi.org/10.1016/j.nuclphysa.2007.01.086} {\bibfield
  {journal} {\bibinfo  {journal} {Nucl. Phys. A}\ }\textbf {\bibinfo {volume}
  {786}},\ \bibinfo {pages} {107} (\bibinfo {year} {2007})},\ \Eprint
  {https://arxiv.org/abs/hep-ph/0610416} {arXiv:hep-ph/0610416} \BibitemShut
  {NoStop}%
\bibitem [{\citenamefont {Marshall}(1963)}]{MAR1}%
  \BibitemOpen
  \bibfield  {author} {\bibinfo {author} {\bibfnamefont {T.~W.}\ \bibnamefont
  {Marshall}},\ }\bibfield  {title} {\bibinfo {title} {Random
  electrodynamics},\ }\href {https://doi.org/10.1098/rspa.1963.0220} {\bibfield
   {journal} {\bibinfo  {journal} {Proc. R. Soc. Lond. A}\ }\textbf {\bibinfo
  {volume} {276}},\ \bibinfo {pages} {475} (\bibinfo {year}
  {1963})}\BibitemShut {NoStop}%
\bibitem [{\citenamefont {Boyer}(1969{\natexlab{a}})}]{BOY1}%
  \BibitemOpen
  \bibfield  {author} {\bibinfo {author} {\bibfnamefont {T.~H.}\ \bibnamefont
  {Boyer}},\ }\bibfield  {title} {\bibinfo {title} {Derivation of the blackbody
  radiation spectrum without quantum assumptions},\ }\href
  {https://doi.org/10.1103/PhysRev.182.1374} {\bibfield  {journal} {\bibinfo
  {journal} {Phys. Rev.}\ }\textbf {\bibinfo {volume} {182}},\ \bibinfo {pages}
  {1374} (\bibinfo {year} {1969}{\natexlab{a}})}\BibitemShut {NoStop}%
\bibitem [{\citenamefont {Boyer}(1969{\natexlab{b}})}]{BOY2}%
  \BibitemOpen
  \bibfield  {author} {\bibinfo {author} {\bibfnamefont {T.~H.}\ \bibnamefont
  {Boyer}},\ }\bibfield  {title} {\bibinfo {title} {Classical statistical
  thermodynamics and electromagnetic zero-point radiation},\ }\href
  {https://doi.org/10.1103/PhysRev.186.1304} {\bibfield  {journal} {\bibinfo
  {journal} {Phys. Rev.}\ }\textbf {\bibinfo {volume} {186}},\ \bibinfo {pages}
  {1304} (\bibinfo {year} {1969}{\natexlab{b}})}\BibitemShut {NoStop}%
\bibitem [{\citenamefont {Boyer}(1975)}]{BOY3}%
  \BibitemOpen
  \bibfield  {author} {\bibinfo {author} {\bibfnamefont {T.~H.}\ \bibnamefont
  {Boyer}},\ }\bibfield  {title} {\bibinfo {title} {Random electrodynamics: The
  theory of classical electrodynamics with classical electromagnetic zero-point
  radiation},\ }\href {https://doi.org/10.1103/PhysRevD.11.790} {\bibfield
  {journal} {\bibinfo  {journal} {Phys. Rev. D}\ }\textbf {\bibinfo {volume}
  {11}},\ \bibinfo {pages} {790} (\bibinfo {year} {1975})}\BibitemShut
  {NoStop}%
\bibitem [{\citenamefont {de~la Pe{\~n}a}\ and\ \citenamefont
  {Cetto}(1996)}]{QDICE}%
  \BibitemOpen
  \bibfield  {author} {\bibinfo {author} {\bibfnamefont {L.}~\bibnamefont
  {de~la Pe{\~n}a}}\ and\ \bibinfo {author} {\bibfnamefont {A.~M.}\
  \bibnamefont {Cetto}},\ }\href@noop {} {\emph {\bibinfo {title} {The Quantum
  Dice: An Introduction to Stochastic Electrodynamics}}}\ (\bibinfo
  {publisher} {Kluwer Academic Publishers},\ \bibinfo {address} {Dordrecht},\
  \bibinfo {year} {1996})\BibitemShut {NoStop}%
\bibitem [{\citenamefont {Wetterich}(2026{\natexlab{a}})}]{CWQF}%
  \BibitemOpen
  \bibfield  {author} {\bibinfo {author} {\bibfnamefont {C.}~\bibnamefont
  {Wetterich}},\ }\bibfield  {title} {\bibinfo {title} {Quantum field theory
  for classical fields},\ }\href {https://doi.org/10.1103/wnpq-hbpq} {\bibfield
   {journal} {\bibinfo  {journal} {APS Open Sci.}\ }\textbf {\bibinfo {volume}
  {1}},\ \bibinfo {pages} {L000179} (\bibinfo {year} {2026}{\natexlab{a}})},\
  \Eprint {https://arxiv.org/abs/2603.05061} {arXiv:2603.05061 [quant-ph]}
  \BibitemShut {NoStop}%
\bibitem [{\citenamefont {Wetterich}(1997{\natexlab{a}})}]{CWET1}%
  \BibitemOpen
  \bibfield  {author} {\bibinfo {author} {\bibfnamefont {C.}~\bibnamefont
  {Wetterich}},\ }\bibfield  {title} {\bibinfo {title} {Time evolution of
  non-equilibrium effective action},\ }\href
  {https://doi.org/10.1103/PhysRevLett.78.3598} {\bibfield  {journal} {\bibinfo
   {journal} {Phys. Rev. Lett.}\ }\textbf {\bibinfo {volume} {78}},\ \bibinfo
  {pages} {3598} (\bibinfo {year} {1997}{\natexlab{a}})},\ \Eprint
  {https://arxiv.org/abs/hep-th/9612206} {arXiv:hep-th/9612206 [hep-th]}
  \BibitemShut {NoStop}%
\bibitem [{\citenamefont {Wetterich}(1997{\natexlab{b}})}]{CWET2}%
  \BibitemOpen
  \bibfield  {author} {\bibinfo {author} {\bibfnamefont {C.}~\bibnamefont
  {Wetterich}},\ }\bibfield  {title} {\bibinfo {title} {Non-equilibrium time
  evolution in quantum field theory},\ }\href
  {https://doi.org/10.1103/PhysRevE.56.2687} {\bibfield  {journal} {\bibinfo
  {journal} {Phys. Rev. E}\ }\textbf {\bibinfo {volume} {56}},\ \bibinfo
  {pages} {2687} (\bibinfo {year} {1997}{\natexlab{b}})},\ \Eprint
  {https://arxiv.org/abs/hep-th/9703006} {arXiv:hep-th/9703006 [hep-th]}
  \BibitemShut {NoStop}%
\bibitem [{\citenamefont {Berges}(2015)}]{BER}%
  \BibitemOpen
  \bibfield  {author} {\bibinfo {author} {\bibfnamefont {J.}~\bibnamefont
  {Berges}},\ }\href@noop {} {\bibinfo {title} {Nonequilibrium quantum fields:
  From cold atoms to cosmology}} (\bibinfo {year} {2015}),\ \Eprint
  {https://arxiv.org/abs/1503.02907} {arXiv:1503.02907 [hep-ph]} \BibitemShut
  {NoStop}%
\bibitem [{\citenamefont {Kamenev}(2011)}]{KAM}%
  \BibitemOpen
  \bibfield  {author} {\bibinfo {author} {\bibfnamefont {A.}~\bibnamefont
  {Kamenev}},\ }\href {https://doi.org/10.1017/CBO9781139003667} {\emph
  {\bibinfo {title} {Field Theory of Non-Equilibrium Systems}}}\ (\bibinfo
  {publisher} {Cambridge University Press},\ \bibinfo {address} {Cambridge},\
  \bibinfo {year} {2011})\BibitemShut {NoStop}%
\bibitem [{\citenamefont {Aarts}\ and\ \citenamefont {Smit}(1998)}]{AS2}%
  \BibitemOpen
  \bibfield  {author} {\bibinfo {author} {\bibfnamefont {G.}~\bibnamefont
  {Aarts}}\ and\ \bibinfo {author} {\bibfnamefont {J.}~\bibnamefont {Smit}},\
  }\bibfield  {title} {\bibinfo {title} {Classical approximation for
  time-dependent quantum field theory: Diagrammatic analysis for hot scalar
  fields},\ }\href {https://doi.org/10.1016/S0550-3213(97)00723-2} {\bibfield
  {journal} {\bibinfo  {journal} {Nucl. Phys. B}\ }\textbf {\bibinfo {volume}
  {511}},\ \bibinfo {pages} {451} (\bibinfo {year} {1998})},\ \Eprint
  {https://arxiv.org/abs/hep-ph/9707342} {arXiv:hep-ph/9707342} \BibitemShut
  {NoStop}%
\bibitem [{\citenamefont {Cooper}\ \emph {et~al.}(2001)\citenamefont {Cooper},
  \citenamefont {Khare},\ and\ \citenamefont {Rose}}]{CKH}%
  \BibitemOpen
  \bibfield  {author} {\bibinfo {author} {\bibfnamefont {F.}~\bibnamefont
  {Cooper}}, \bibinfo {author} {\bibfnamefont {A.}~\bibnamefont {Khare}},\ and\
  \bibinfo {author} {\bibfnamefont {H.}~\bibnamefont {Rose}},\ }\bibfield
  {title} {\bibinfo {title} {Classical limit of time-dependent quantum field
  theory: A {S}chwinger--{D}yson approach},\ }\href
  {https://doi.org/10.1016/S0370-2693(01)00872-3} {\bibfield  {journal}
  {\bibinfo  {journal} {Phys. Lett. B}\ }\textbf {\bibinfo {volume} {515}},\
  \bibinfo {pages} {463} (\bibinfo {year} {2001})},\ \Eprint
  {https://arxiv.org/abs/hep-ph/0106113} {arXiv:hep-ph/0106113} \BibitemShut
  {NoStop}%
\bibitem [{\citenamefont {Epelbaum}\ \emph {et~al.}(2014)\citenamefont
  {Epelbaum}, \citenamefont {Gelis},\ and\ \citenamefont {Wu}}]{EGW}%
  \BibitemOpen
  \bibfield  {author} {\bibinfo {author} {\bibfnamefont {T.}~\bibnamefont
  {Epelbaum}}, \bibinfo {author} {\bibfnamefont {F.}~\bibnamefont {Gelis}},\
  and\ \bibinfo {author} {\bibfnamefont {B.}~\bibnamefont {Wu}},\ }\bibfield
  {title} {\bibinfo {title} {Nonrenormalizability of the classical statistical
  approximation},\ }\href {https://doi.org/10.1103/PhysRevD.90.065029}
  {\bibfield  {journal} {\bibinfo  {journal} {Phys. Rev. D}\ }\textbf {\bibinfo
  {volume} {90}},\ \bibinfo {pages} {065029} (\bibinfo {year} {2014})},\
  \Eprint {https://arxiv.org/abs/1402.0115} {arXiv:1402.0115 [hep-ph]}
  \BibitemShut {NoStop}%
\bibitem [{\citenamefont {Aarts}\ and\ \citenamefont {Smit}(1997)}]{AS1}%
  \BibitemOpen
  \bibfield  {author} {\bibinfo {author} {\bibfnamefont {G.}~\bibnamefont
  {Aarts}}\ and\ \bibinfo {author} {\bibfnamefont {J.}~\bibnamefont {Smit}},\
  }\bibfield  {title} {\bibinfo {title} {Finiteness of hot classical scalar
  field theory and the plasmon damping rate},\ }\href
  {https://doi.org/10.1016/S0370-2693(96)01624-3} {\bibfield  {journal}
  {\bibinfo  {journal} {Phys. Lett. B}\ }\textbf {\bibinfo {volume} {393}},\
  \bibinfo {pages} {395} (\bibinfo {year} {1997})},\ \Eprint
  {https://arxiv.org/abs/hep-ph/9610415} {arXiv:hep-ph/9610415} \BibitemShut
  {NoStop}%
\bibitem [{\citenamefont {Buchm{\"u}ller}\ and\ \citenamefont
  {Jakov{\'a}c}(1997)}]{BUJA}%
  \BibitemOpen
  \bibfield  {author} {\bibinfo {author} {\bibfnamefont {W.}~\bibnamefont
  {Buchm{\"u}ller}}\ and\ \bibinfo {author} {\bibfnamefont {A.}~\bibnamefont
  {Jakov{\'a}c}},\ }\bibfield  {title} {\bibinfo {title} {Classical statistical
  mechanics and {L}andau damping},\ }\href
  {https://doi.org/10.1016/S0370-2693(97)00746-6} {\bibfield  {journal}
  {\bibinfo  {journal} {Phys. Lett. B}\ }\textbf {\bibinfo {volume} {407}},\
  \bibinfo {pages} {39} (\bibinfo {year} {1997})},\ \Eprint
  {https://arxiv.org/abs/hep-ph/9705452} {arXiv:hep-ph/9705452} \BibitemShut
  {NoStop}%
\bibitem [{\citenamefont {Berges}\ \emph {et~al.}(2012)\citenamefont {Berges},
  \citenamefont {Blaizot},\ and\ \citenamefont {Gelis}}]{BBG}%
  \BibitemOpen
  \bibfield  {author} {\bibinfo {author} {\bibfnamefont {J.}~\bibnamefont
  {Berges}}, \bibinfo {author} {\bibfnamefont {J.-P.}\ \bibnamefont
  {Blaizot}},\ and\ \bibinfo {author} {\bibfnamefont {F.}~\bibnamefont
  {Gelis}},\ }\bibfield  {title} {\bibinfo {title} {{EMMI} rapid reaction task
  force on ``{T}hermalization in non-{A}belian plasmas''},\ }\href
  {https://doi.org/10.1088/0954-3899/39/8/085115} {\bibfield  {journal}
  {\bibinfo  {journal} {J. Phys. G}\ }\textbf {\bibinfo {volume} {39}},\
  \bibinfo {pages} {085115} (\bibinfo {year} {2012})},\ \Eprint
  {https://arxiv.org/abs/1203.2042} {arXiv:1203.2042 [hep-ph]} \BibitemShut
  {NoStop}%
\bibitem [{\citenamefont {Wetterich}(2026{\natexlab{b}})}]{CWDS}%
  \BibitemOpen
  \bibfield  {author} {\bibinfo {author} {\bibfnamefont {C.}~\bibnamefont
  {Wetterich}},\ }\href@noop {} {\bibinfo {title} {Classical probabilistic
  realisation of quantum double-slit interference}} (\bibinfo {year}
  {2026}{\natexlab{b}}),\ \Eprint {https://arxiv.org/abs/2607.11580}
  {arXiv:2607.11580 [quant-ph]} \BibitemShut {NoStop}%
\bibitem [{\citenamefont {Wetterich}(2012)}]{CWQP1}%
  \BibitemOpen
  \bibfield  {author} {\bibinfo {author} {\bibfnamefont {C.}~\bibnamefont
  {Wetterich}},\ }\bibfield  {title} {\bibinfo {title} {Quantum particles from
  classical probabilities in phase space},\ }\href@noop {} {\bibfield
  {journal} {\bibinfo  {journal} {Int. J. Theor. Phys.}\ }\textbf {\bibinfo
  {volume} {51}},\ \bibinfo {pages} {3236} (\bibinfo {year} {2012})},\ \Eprint
  {https://arxiv.org/abs/1003.0772} {arXiv:1003.0772 [quant-ph]} \BibitemShut
  {NoStop}%
\bibitem [{\citenamefont {Wetterich}(2018{\natexlab{a}})}]{CWIT}%
  \BibitemOpen
  \bibfield  {author} {\bibinfo {author} {\bibfnamefont {C.}~\bibnamefont
  {Wetterich}},\ }\bibfield  {title} {\bibinfo {title} {Information transport
  in classical statistical systems},\ }\href
  {https://doi.org/10.1016/j.nuclphysb.2017.12.008} {\bibfield  {journal}
  {\bibinfo  {journal} {Nucl. Phys. B}\ }\textbf {\bibinfo {volume} {927}},\
  \bibinfo {pages} {35} (\bibinfo {year} {2018}{\natexlab{a}})},\ \Eprint
  {https://arxiv.org/abs/1611.04820} {arXiv:1611.04820 [cond-mat.stat-mech]}
  \BibitemShut {NoStop}%
\bibitem [{\citenamefont {Wetterich}(2018{\natexlab{b}})}]{CWQFC}%
  \BibitemOpen
  \bibfield  {author} {\bibinfo {author} {\bibfnamefont {C.}~\bibnamefont
  {Wetterich}},\ }\bibfield  {title} {\bibinfo {title} {Quantum formalism for
  classical statistics},\ }\href {https://doi.org/10.1016/j.aop.2018.03.022}
  {\bibfield  {journal} {\bibinfo  {journal} {Ann. Phys.}\ }\textbf {\bibinfo
  {volume} {393}},\ \bibinfo {pages} {1} (\bibinfo {year}
  {2018}{\natexlab{b}})},\ \Eprint {https://arxiv.org/abs/1706.01772}
  {arXiv:1706.01772 [quant-ph]} \BibitemShut {NoStop}%
\bibitem [{\citenamefont {Koopman}(1931)}]{KOOP}%
  \BibitemOpen
  \bibfield  {author} {\bibinfo {author} {\bibfnamefont {B.~O.}\ \bibnamefont
  {Koopman}},\ }\bibfield  {title} {\bibinfo {title} {Hamiltonian systems and
  transformation in {H}ilbert space},\ }\href
  {https://doi.org/10.1073/pnas.17.5.315} {\bibfield  {journal} {\bibinfo
  {journal} {Proc. Natl. Acad. Sci. USA}\ }\textbf {\bibinfo {volume} {17}},\
  \bibinfo {pages} {315} (\bibinfo {year} {1931})}\BibitemShut {NoStop}%
\bibitem [{\citenamefont {von Neumann}(1932)}]{VNEU}%
  \BibitemOpen
  \bibfield  {author} {\bibinfo {author} {\bibfnamefont {J.}~\bibnamefont {von
  Neumann}},\ }\bibfield  {title} {\bibinfo {title} {Zur {O}peratorenmethode in
  der klassischen {M}echanik},\ }\href {https://doi.org/10.2307/1968537}
  {\bibfield  {journal} {\bibinfo  {journal} {Ann. Math.}\ }\textbf {\bibinfo
  {volume} {33}},\ \bibinfo {pages} {587} (\bibinfo {year} {1932})}\BibitemShut
  {NoStop}%
\bibitem [{\citenamefont {Sch{\"o}nberg}(1952)}]{SCHB}%
  \BibitemOpen
  \bibfield  {author} {\bibinfo {author} {\bibfnamefont {M.}~\bibnamefont
  {Sch{\"o}nberg}},\ }\bibfield  {title} {\bibinfo {title} {Application of
  second quantization methods to the classical statistical mechanics},\ }\href
  {https://doi.org/10.1007/BF02782925} {\bibfield  {journal} {\bibinfo
  {journal} {Nuovo Cimento}\ }\textbf {\bibinfo {volume} {9}},\ \bibinfo
  {pages} {1139} (\bibinfo {year} {1952})}\BibitemShut {NoStop}%
\bibitem [{\citenamefont {Kandrup}(1998)}]{KAN}%
  \BibitemOpen
  \bibfield  {author} {\bibinfo {author} {\bibfnamefont {H.~E.}\ \bibnamefont
  {Kandrup}},\ }\bibfield  {title} {\bibinfo {title} {Phase mixing in
  time-independent {H}amiltonian systems},\ }\href
  {https://doi.org/10.1046/j.1365-8711.1998.02063.x} {\bibfield  {journal}
  {\bibinfo  {journal} {Mon. Not. R. Astron. Soc.}\ }\textbf {\bibinfo {volume}
  {301}},\ \bibinfo {pages} {960} (\bibinfo {year} {1998})},\ \Eprint
  {https://arxiv.org/abs/astro-ph/9809100} {arXiv:astro-ph/9809100}
  \BibitemShut {NoStop}%
\bibitem [{\citenamefont {Man'ko}\ and\ \citenamefont {Marmo}(1999)}]{MAMA}%
  \BibitemOpen
  \bibfield  {author} {\bibinfo {author} {\bibfnamefont {V.~I.}\ \bibnamefont
  {Man'ko}}\ and\ \bibinfo {author} {\bibfnamefont {G.}~\bibnamefont {Marmo}},\
  }\bibfield  {title} {\bibinfo {title} {Probability distributions and
  {H}ilbert spaces: quantum and classical systems},\ }\href
  {https://doi.org/10.1238/Physica.Regular.060a00111} {\bibfield  {journal}
  {\bibinfo  {journal} {Phys. Scr.}\ }\textbf {\bibinfo {volume} {60}},\
  \bibinfo {pages} {111} (\bibinfo {year} {1999})},\ \Eprint
  {https://arxiv.org/abs/quant-ph/9903021} {arXiv:quant-ph/9903021 [quant-ph]}
  \BibitemShut {NoStop}%
\bibitem [{\citenamefont {Mauro}(2002)}]{MAU}%
  \BibitemOpen
  \bibfield  {author} {\bibinfo {author} {\bibfnamefont {D.}~\bibnamefont
  {Mauro}},\ }\bibfield  {title} {\bibinfo {title} {On {K}oopman--von {N}eumann
  waves},\ }\href {https://doi.org/10.1142/S0217751X02009680} {\bibfield
  {journal} {\bibinfo  {journal} {Int. J. Mod. Phys. A}\ }\textbf {\bibinfo
  {volume} {17}},\ \bibinfo {pages} {1301} (\bibinfo {year} {2002})},\ \Eprint
  {https://arxiv.org/abs/quant-ph/0105112} {arXiv:quant-ph/0105112 [quant-ph]}
  \BibitemShut {NoStop}%
\bibitem [{\citenamefont {Gozzi}\ and\ \citenamefont {Mauro}(2004)}]{GOMA}%
  \BibitemOpen
  \bibfield  {author} {\bibinfo {author} {\bibfnamefont {E.}~\bibnamefont
  {Gozzi}}\ and\ \bibinfo {author} {\bibfnamefont {D.}~\bibnamefont {Mauro}},\
  }\bibfield  {title} {\bibinfo {title} {On {K}oopman--von {N}eumann waves
  {II}},\ }\href {https://doi.org/10.1142/S0217751X04017872} {\bibfield
  {journal} {\bibinfo  {journal} {Int. J. Mod. Phys. A}\ }\textbf {\bibinfo
  {volume} {19}},\ \bibinfo {pages} {1475} (\bibinfo {year} {2004})},\ \Eprint
  {https://arxiv.org/abs/quant-ph/0306029} {arXiv:quant-ph/0306029 [quant-ph]}
  \BibitemShut {NoStop}%
\bibitem [{\citenamefont {Nikoli{\'c}}(2006)}]{NIC}%
  \BibitemOpen
  \bibfield  {author} {\bibinfo {author} {\bibfnamefont {H.}~\bibnamefont
  {Nikoli{\'c}}},\ }\bibfield  {title} {\bibinfo {title} {Classical mechanics
  without determinism},\ }\href {https://doi.org/10.1007/s10702-006-1009-2}
  {\bibfield  {journal} {\bibinfo  {journal} {Found. Phys. Lett.}\ }\textbf
  {\bibinfo {volume} {19}},\ \bibinfo {pages} {553} (\bibinfo {year} {2006})},\
  \Eprint {https://arxiv.org/abs/quant-ph/0505143} {arXiv:quant-ph/0505143
  [quant-ph]} \BibitemShut {NoStop}%
\bibitem [{\citenamefont {Nikoli{\'c}}(2007)}]{NAKL}%
  \BibitemOpen
  \bibfield  {author} {\bibinfo {author} {\bibfnamefont {H.}~\bibnamefont
  {Nikoli{\'c}}},\ }\bibfield  {title} {\bibinfo {title} {Classical mechanics
  as nonlinear quantum mechanics},\ }\href {https://doi.org/10.1063/1.2827300}
  {\bibfield  {journal} {\bibinfo  {journal} {AIP Conf. Proc.}\ }\textbf
  {\bibinfo {volume} {962}},\ \bibinfo {pages} {162} (\bibinfo {year}
  {2007})},\ \Eprint {https://arxiv.org/abs/0707.2319} {arXiv:0707.2319
  [quant-ph]} \BibitemShut {NoStop}%
\bibitem [{\citenamefont {Volovich}(2011)}]{VOLO}%
  \BibitemOpen
  \bibfield  {author} {\bibinfo {author} {\bibfnamefont {I.~V.}\ \bibnamefont
  {Volovich}},\ }\bibfield  {title} {\bibinfo {title} {Randomness in classical
  mechanics and quantum mechanics},\ }\href
  {https://doi.org/10.1007/s10701-010-9450-2} {\bibfield  {journal} {\bibinfo
  {journal} {Found. Phys.}\ }\textbf {\bibinfo {volume} {41}},\ \bibinfo
  {pages} {516} (\bibinfo {year} {2011})},\ \Eprint
  {https://arxiv.org/abs/0910.5391} {arXiv:0910.5391 [quant-ph]} \BibitemShut
  {NoStop}%
\bibitem [{\citenamefont {Bondar}\ \emph {et~al.}(2012)\citenamefont {Bondar},
  \citenamefont {Cabrera}, \citenamefont {Lompay}, \citenamefont {Ivanov},\
  and\ \citenamefont {Rabitz}}]{BON}%
  \BibitemOpen
  \bibfield  {author} {\bibinfo {author} {\bibfnamefont {D.~I.}\ \bibnamefont
  {Bondar}}, \bibinfo {author} {\bibfnamefont {R.}~\bibnamefont {Cabrera}},
  \bibinfo {author} {\bibfnamefont {R.~R.}\ \bibnamefont {Lompay}}, \bibinfo
  {author} {\bibfnamefont {M.~Y.}\ \bibnamefont {Ivanov}},\ and\ \bibinfo
  {author} {\bibfnamefont {H.~A.}\ \bibnamefont {Rabitz}},\ }\bibfield  {title}
  {\bibinfo {title} {Operational dynamic modeling transcending quantum and
  classical mechanics},\ }\href
  {https://doi.org/10.1103/PhysRevLett.109.190403} {\bibfield  {journal}
  {\bibinfo  {journal} {Phys. Rev. Lett.}\ }\textbf {\bibinfo {volume} {109}},\
  \bibinfo {pages} {190403} (\bibinfo {year} {2012})},\ \Eprint
  {https://arxiv.org/abs/1105.4014} {arXiv:1105.4014 [quant-ph]} \BibitemShut
  {NoStop}%
\bibitem [{\citenamefont {Arbabi}\ and\ \citenamefont {Mezi{\'c}}(2017)}]{AME}%
  \BibitemOpen
  \bibfield  {author} {\bibinfo {author} {\bibfnamefont {H.}~\bibnamefont
  {Arbabi}}\ and\ \bibinfo {author} {\bibfnamefont {I.}~\bibnamefont
  {Mezi{\'c}}},\ }\bibfield  {title} {\bibinfo {title} {Ergodic theory, dynamic
  mode decomposition, and computation of spectral properties of the {K}oopman
  operator},\ }\href@noop {} {\bibfield  {journal} {\bibinfo  {journal} {SIAM
  J. Appl. Dyn. Syst.}\ }\textbf {\bibinfo {volume} {16}},\ \bibinfo {pages}
  {2096} (\bibinfo {year} {2017})},\ \Eprint {https://arxiv.org/abs/1611.06664}
  {arXiv:1611.06664 [math.DS]} \BibitemShut {NoStop}%
\bibitem [{\citenamefont {Korda}\ \emph {et~al.}(2020)\citenamefont {Korda},
  \citenamefont {Putinar},\ and\ \citenamefont {Mezi{\'c}}}]{KPM}%
  \BibitemOpen
  \bibfield  {author} {\bibinfo {author} {\bibfnamefont {M.}~\bibnamefont
  {Korda}}, \bibinfo {author} {\bibfnamefont {M.}~\bibnamefont {Putinar}},\
  and\ \bibinfo {author} {\bibfnamefont {I.}~\bibnamefont {Mezi{\'c}}},\
  }\bibfield  {title} {\bibinfo {title} {Data-driven spectral analysis of the
  {K}oopman operator},\ }\href {https://doi.org/10.1016/j.acha.2018.08.002}
  {\bibfield  {journal} {\bibinfo  {journal} {Appl. Comput. Harmon. Anal.}\
  }\textbf {\bibinfo {volume} {48}},\ \bibinfo {pages} {599} (\bibinfo {year}
  {2020})},\ \Eprint {https://arxiv.org/abs/1710.06532} {arXiv:1710.06532
  [math.DS]} \BibitemShut {NoStop}%
\bibitem [{\citenamefont {Arsiwalla}\ \emph {et~al.}(2023)\citenamefont
  {Arsiwalla}, \citenamefont {Chester},\ and\ \citenamefont {Kauffman}}]{ACKA}%
  \BibitemOpen
  \bibfield  {author} {\bibinfo {author} {\bibfnamefont {X.~D.}\ \bibnamefont
  {Arsiwalla}}, \bibinfo {author} {\bibfnamefont {D.}~\bibnamefont {Chester}},\
  and\ \bibinfo {author} {\bibfnamefont {L.~H.}\ \bibnamefont {Kauffman}},\
  }\bibfield  {title} {\bibinfo {title} {On the operator origins of classical
  and quantum wave functions},\ }\href
  {https://doi.org/10.1007/s40509-023-00311-6} {\bibfield  {journal} {\bibinfo
  {journal} {Quantum Stud.: Math. Found.}\ }\textbf {\bibinfo {volume} {11}},\
  \bibinfo {pages} {193} (\bibinfo {year} {2023})},\ \Eprint
  {https://arxiv.org/abs/2211.01838} {arXiv:2211.01838 [math-ph]} \BibitemShut
  {NoStop}%
\bibitem [{\citenamefont {Chru{\'s}ci{\'n}ski}(2006)}]{CHRU}%
  \BibitemOpen
  \bibfield  {author} {\bibinfo {author} {\bibfnamefont {D.}~\bibnamefont
  {Chru{\'s}ci{\'n}ski}},\ }\bibfield  {title} {\bibinfo {title} {Koopman's
  approach to dissipation},\ }\href
  {https://doi.org/10.1016/S0034-4877(06)80023-6} {\bibfield  {journal}
  {\bibinfo  {journal} {Rep. Math. Phys.}\ }\textbf {\bibinfo {volume} {57}},\
  \bibinfo {pages} {319} (\bibinfo {year} {2006})}\BibitemShut {NoStop}%
\bibitem [{\citenamefont {Klein}(2018)}]{KLE}%
  \BibitemOpen
  \bibfield  {author} {\bibinfo {author} {\bibfnamefont {U.}~\bibnamefont
  {Klein}},\ }\bibfield  {title} {\bibinfo {title} {From {K}oopman--von
  {N}eumann theory to quantum theory},\ }\href
  {https://doi.org/10.1007/s40509-017-0113-2} {\bibfield  {journal} {\bibinfo
  {journal} {Quantum Stud.: Math. Found.}\ }\textbf {\bibinfo {volume} {5}},\
  \bibinfo {pages} {219} (\bibinfo {year} {2018})},\ \Eprint
  {https://arxiv.org/abs/1705.07427} {arXiv:1705.07427 [quant-ph]} \BibitemShut
  {NoStop}%
\bibitem [{\citenamefont {Mezi{\'c}}(2022)}]{MEZI}%
  \BibitemOpen
  \bibfield  {author} {\bibinfo {author} {\bibfnamefont {I.}~\bibnamefont
  {Mezi{\'c}}},\ }\bibfield  {title} {\bibinfo {title} {On numerical
  approximations of the {K}oopman operator},\ }\href
  {https://doi.org/10.3390/math10071180} {\bibfield  {journal} {\bibinfo
  {journal} {Mathematics}\ }\textbf {\bibinfo {volume} {10}},\ \bibinfo {pages}
  {1180} (\bibinfo {year} {2022})},\ \Eprint {https://arxiv.org/abs/2009.05883}
  {arXiv:2009.05883 [math.DS]} \BibitemShut {NoStop}%
\bibitem [{\citenamefont {Darling}\ and\ \citenamefont {Widrow}(2024)}]{DW}%
  \BibitemOpen
  \bibfield  {author} {\bibinfo {author} {\bibfnamefont {K.}~\bibnamefont
  {Darling}}\ and\ \bibinfo {author} {\bibfnamefont {L.~M.}\ \bibnamefont
  {Widrow}},\ }\bibfield  {title} {\bibinfo {title} {Linear operator theory of
  phase mixing},\ }\href {https://doi.org/10.1093/mnras/stae1775} {\bibfield
  {journal} {\bibinfo  {journal} {Mon. Not. R. Astron. Soc.}\ }\textbf
  {\bibinfo {volume} {533}},\ \bibinfo {pages} {79} (\bibinfo {year}
  {2024})}\BibitemShut {NoStop}%
\bibitem [{\citenamefont {Gozzi}(1988)}]{GOZ1}%
  \BibitemOpen
  \bibfield  {author} {\bibinfo {author} {\bibfnamefont {E.}~\bibnamefont
  {Gozzi}},\ }\bibfield  {title} {\bibinfo {title} {Hidden {BRS} invariance in
  classical mechanics},\ }\href {https://doi.org/10.1016/0370-2693(88)90611-9}
  {\bibfield  {journal} {\bibinfo  {journal} {Phys. Lett. B}\ }\textbf
  {\bibinfo {volume} {201}},\ \bibinfo {pages} {525} (\bibinfo {year}
  {1988})}\BibitemShut {NoStop}%
\bibitem [{\citenamefont {Gozzi}\ \emph {et~al.}(1989)\citenamefont {Gozzi},
  \citenamefont {Reuter},\ and\ \citenamefont {Thacker}}]{GRT}%
  \BibitemOpen
  \bibfield  {author} {\bibinfo {author} {\bibfnamefont {E.}~\bibnamefont
  {Gozzi}}, \bibinfo {author} {\bibfnamefont {M.}~\bibnamefont {Reuter}},\ and\
  \bibinfo {author} {\bibfnamefont {W.~D.}\ \bibnamefont {Thacker}},\
  }\bibfield  {title} {\bibinfo {title} {Hidden {BRS} invariance in classical
  mechanics. {II}},\ }\href {https://doi.org/10.1103/PhysRevD.40.3363}
  {\bibfield  {journal} {\bibinfo  {journal} {Phys. Rev. D}\ }\textbf {\bibinfo
  {volume} {40}},\ \bibinfo {pages} {3363} (\bibinfo {year}
  {1989})}\BibitemShut {NoStop}%
\bibitem [{\citenamefont {Wetterich}(2026{\natexlab{c}})}]{CWQP}%
  \BibitemOpen
  \bibfield  {author} {\bibinfo {author} {\bibfnamefont {C.}~\bibnamefont
  {Wetterich}},\ }\href@noop {} {\bibinfo {title} {Quantum observables for
  probabilistic classical particles}} (\bibinfo {year} {2026}{\natexlab{c}}),\
  \Eprint {https://arxiv.org/abs/2607.13937} {arXiv:2607.13937 [quant-ph]}
  \BibitemShut {NoStop}%
\bibitem [{\citenamefont {Bell}(1964)}]{BEL1}%
  \BibitemOpen
  \bibfield  {author} {\bibinfo {author} {\bibfnamefont {J.~S.}\ \bibnamefont
  {Bell}},\ }\bibfield  {title} {\bibinfo {title} {On the {E}instein {P}odolsky
  {R}osen paradox},\ }\href
  {https://doi.org/10.1103/PhysicsPhysiqueFizika.1.195} {\bibfield  {journal}
  {\bibinfo  {journal} {Physics Physique Fizika}\ }\textbf {\bibinfo {volume}
  {1}},\ \bibinfo {pages} {195} (\bibinfo {year} {1964})}\BibitemShut {NoStop}%
\bibitem [{\citenamefont {Clauser}\ \emph {et~al.}(1969)\citenamefont
  {Clauser}, \citenamefont {Horne}, \citenamefont {Shimony},\ and\
  \citenamefont {Holt}}]{CHSH}%
  \BibitemOpen
  \bibfield  {author} {\bibinfo {author} {\bibfnamefont {J.~F.}\ \bibnamefont
  {Clauser}}, \bibinfo {author} {\bibfnamefont {M.~A.}\ \bibnamefont {Horne}},
  \bibinfo {author} {\bibfnamefont {A.}~\bibnamefont {Shimony}},\ and\ \bibinfo
  {author} {\bibfnamefont {R.~A.}\ \bibnamefont {Holt}},\ }\bibfield  {title}
  {\bibinfo {title} {Proposed experiment to test local hidden-variable
  theories},\ }\href {https://doi.org/10.1103/PhysRevLett.23.880} {\bibfield
  {journal} {\bibinfo  {journal} {Phys. Rev. Lett.}\ }\textbf {\bibinfo
  {volume} {23}},\ \bibinfo {pages} {880} (\bibinfo {year} {1969})}\BibitemShut
  {NoStop}%
\bibitem [{\citenamefont {Husimi}(1940)}]{HUS}%
  \BibitemOpen
  \bibfield  {author} {\bibinfo {author} {\bibfnamefont {K.}~\bibnamefont
  {Husimi}},\ }\bibfield  {title} {\bibinfo {title} {Some formal properties of
  the density matrix},\ }\href@noop {} {\bibfield  {journal} {\bibinfo
  {journal} {Proc. Phys.-Math. Soc. Japan}\ }\textbf {\bibinfo {volume} {22}},\
  \bibinfo {pages} {264} (\bibinfo {year} {1940})}\BibitemShut {NoStop}%
\bibitem [{\citenamefont {Glauber}(1963{\natexlab{a}})}]{GLA}%
  \BibitemOpen
  \bibfield  {author} {\bibinfo {author} {\bibfnamefont {R.~J.}\ \bibnamefont
  {Glauber}},\ }\bibfield  {title} {\bibinfo {title} {Coherent and incoherent
  states of the radiation field},\ }\href
  {https://doi.org/10.1103/PhysRev.131.2766} {\bibfield  {journal} {\bibinfo
  {journal} {Phys. Rev.}\ }\textbf {\bibinfo {volume} {131}},\ \bibinfo {pages}
  {2766} (\bibinfo {year} {1963}{\natexlab{a}})}\BibitemShut {NoStop}%
\bibitem [{\citenamefont {Sudarshan}(1963)}]{SUD}%
  \BibitemOpen
  \bibfield  {author} {\bibinfo {author} {\bibfnamefont {E.~C.~G.}\
  \bibnamefont {Sudarshan}},\ }\bibfield  {title} {\bibinfo {title}
  {Equivalence of semiclassical and quantum mechanical descriptions of
  statistical light beams},\ }\href
  {https://doi.org/10.1103/PhysRevLett.10.277} {\bibfield  {journal} {\bibinfo
  {journal} {Phys. Rev. Lett.}\ }\textbf {\bibinfo {volume} {10}},\ \bibinfo
  {pages} {277} (\bibinfo {year} {1963})}\BibitemShut {NoStop}%
\bibitem [{\citenamefont {Glauber}(1963{\natexlab{b}})}]{GLA2}%
  \BibitemOpen
  \bibfield  {author} {\bibinfo {author} {\bibfnamefont {R.~J.}\ \bibnamefont
  {Glauber}},\ }\bibfield  {title} {\bibinfo {title} {The quantum theory of
  optical coherence},\ }\href {https://doi.org/10.1103/PhysRev.130.2529}
  {\bibfield  {journal} {\bibinfo  {journal} {Phys. Rev.}\ }\textbf {\bibinfo
  {volume} {130}},\ \bibinfo {pages} {2529} (\bibinfo {year}
  {1963}{\natexlab{b}})}\BibitemShut {NoStop}%
\bibitem [{\citenamefont {Cahill}\ and\ \citenamefont {Glauber}(1969)}]{CAGL}%
  \BibitemOpen
  \bibfield  {author} {\bibinfo {author} {\bibfnamefont {K.~E.}\ \bibnamefont
  {Cahill}}\ and\ \bibinfo {author} {\bibfnamefont {R.~J.}\ \bibnamefont
  {Glauber}},\ }\bibfield  {title} {\bibinfo {title} {Density operators and
  quasiprobability distributions},\ }\href
  {https://doi.org/10.1103/PhysRev.177.1882} {\bibfield  {journal} {\bibinfo
  {journal} {Phys. Rev.}\ }\textbf {\bibinfo {volume} {177}},\ \bibinfo {pages}
  {1882} (\bibinfo {year} {1969})}\BibitemShut {NoStop}%
\bibitem [{\citenamefont {Wigner}(1932)}]{WIG}%
  \BibitemOpen
  \bibfield  {author} {\bibinfo {author} {\bibfnamefont {E.}~\bibnamefont
  {Wigner}},\ }\bibfield  {title} {\bibinfo {title} {On the quantum correction
  for thermodynamic equilibrium},\ }\href
  {https://doi.org/10.1103/PhysRev.40.749} {\bibfield  {journal} {\bibinfo
  {journal} {Phys. Rev.}\ }\textbf {\bibinfo {volume} {40}},\ \bibinfo {pages}
  {749} (\bibinfo {year} {1932})}\BibitemShut {NoStop}%
\bibitem [{\citenamefont {Wetterich}(2025)}]{CWPW}%
  \BibitemOpen
  \bibfield  {author} {\bibinfo {author} {\bibfnamefont {C.}~\bibnamefont
  {Wetterich}},\ }\href@noop {} {\emph {\bibinfo {title} {The Probabilistic
  World}}}\ (\bibinfo  {publisher} {Springer Nature},\ \bibinfo {address}
  {Heidelberg},\ \bibinfo {year} {2025})\ \Eprint
  {https://arxiv.org/abs/2011.02867} {arXiv:2011.02867 [quant-ph]} \BibitemShut
  {NoStop}%
\bibitem [{\citenamefont {Bonini}\ and\ \citenamefont
  {Wetterich}(1999)}]{BOWE}%
  \BibitemOpen
  \bibfield  {author} {\bibinfo {author} {\bibfnamefont {G.~F.}\ \bibnamefont
  {Bonini}}\ and\ \bibinfo {author} {\bibfnamefont {C.}~\bibnamefont
  {Wetterich}},\ }\bibfield  {title} {\bibinfo {title} {Time evolution of
  correlation functions and thermalization},\ }\href
  {https://doi.org/10.1103/PhysRevD.60.105026} {\bibfield  {journal} {\bibinfo
  {journal} {Phys. Rev. D}\ }\textbf {\bibinfo {volume} {60}},\ \bibinfo
  {pages} {105026} (\bibinfo {year} {1999})},\ \Eprint
  {https://arxiv.org/abs/hep-ph/9907533} {arXiv:hep-ph/9907533 [hep-ph]}
  \BibitemShut {NoStop}%
\bibitem [{\citenamefont {Aarts}\ \emph {et~al.}(2000)\citenamefont {Aarts},
  \citenamefont {Bonini},\ and\ \citenamefont {Wetterich}}]{ABW}%
  \BibitemOpen
  \bibfield  {author} {\bibinfo {author} {\bibfnamefont {G.}~\bibnamefont
  {Aarts}}, \bibinfo {author} {\bibfnamefont {G.~F.}\ \bibnamefont {Bonini}},\
  and\ \bibinfo {author} {\bibfnamefont {C.}~\bibnamefont {Wetterich}},\
  }\bibfield  {title} {\bibinfo {title} {On thermalization in classical scalar
  field theory},\ }\href {https://doi.org/10.1016/S0550-3213(00)00447-8}
  {\bibfield  {journal} {\bibinfo  {journal} {Nucl. Phys. B}\ }\textbf
  {\bibinfo {volume} {587}},\ \bibinfo {pages} {403} (\bibinfo {year}
  {2000})},\ \Eprint {https://arxiv.org/abs/hep-ph/0003262}
  {arXiv:hep-ph/0003262} \BibitemShut {NoStop}%
\bibitem [{\citenamefont {Zache}\ \emph {et~al.}(2020)\citenamefont {Zache},
  \citenamefont {Schweigler}, \citenamefont {Erne}, \citenamefont
  {Schmiedmayer},\ and\ \citenamefont {Berges}}]{TSE}%
  \BibitemOpen
  \bibfield  {author} {\bibinfo {author} {\bibfnamefont {T.~V.}\ \bibnamefont
  {Zache}}, \bibinfo {author} {\bibfnamefont {T.}~\bibnamefont {Schweigler}},
  \bibinfo {author} {\bibfnamefont {S.}~\bibnamefont {Erne}}, \bibinfo {author}
  {\bibfnamefont {J.}~\bibnamefont {Schmiedmayer}},\ and\ \bibinfo {author}
  {\bibfnamefont {J.}~\bibnamefont {Berges}},\ }\bibfield  {title} {\bibinfo
  {title} {Extracting the field theory description of a quantum many-body
  system from experimental data},\ }\href
  {https://doi.org/10.1103/PhysRevX.10.011020} {\bibfield  {journal} {\bibinfo
  {journal} {Phys. Rev. X}\ }\textbf {\bibinfo {volume} {10}},\ \bibinfo
  {pages} {011020} (\bibinfo {year} {2020})},\ \Eprint
  {https://arxiv.org/abs/1909.12815} {arXiv:1909.12815 [cond-mat.quant-gas]}
  \BibitemShut {NoStop}%
\bibitem [{\citenamefont {Ott}\ \emph {et~al.}(2023)\citenamefont {Ott},
  \citenamefont {Zache},\ and\ \citenamefont {Berges}}]{OZB}%
  \BibitemOpen
  \bibfield  {author} {\bibinfo {author} {\bibfnamefont {R.}~\bibnamefont
  {Ott}}, \bibinfo {author} {\bibfnamefont {T.~V.}\ \bibnamefont {Zache}},\
  and\ \bibinfo {author} {\bibfnamefont {J.}~\bibnamefont {Berges}},\
  }\bibfield  {title} {\bibinfo {title} {Equal-time approach to real-time
  dynamics of quantum fields},\ }\href
  {https://doi.org/10.21468/SciPostPhys.14.2.011} {\bibfield  {journal}
  {\bibinfo  {journal} {SciPost Phys.}\ }\textbf {\bibinfo {volume} {14}},\
  \bibinfo {pages} {011} (\bibinfo {year} {2023})},\ \Eprint
  {https://arxiv.org/abs/2204.06463} {arXiv:2204.06463 [cond-mat.quant-gas]}
  \BibitemShut {NoStop}%
\bibitem [{\citenamefont {Nachbagauer}(1997)}]{NAC}%
  \BibitemOpen
  \bibfield  {author} {\bibinfo {author} {\bibfnamefont {H.}~\bibnamefont
  {Nachbagauer}},\ }\href@noop {} {\bibinfo {title} {Wigner functionals and
  their dynamics in quantum field theory}} (\bibinfo {year} {1997}),\ \Eprint
  {https://arxiv.org/abs/hep-th/9703105} {arXiv:hep-th/9703105 [hep-th]}
  \BibitemShut {NoStop}%
\bibitem [{\citenamefont {B{\"o}deker}\ \emph {et~al.}(1995)\citenamefont
  {B{\"o}deker}, \citenamefont {McLerran},\ and\ \citenamefont
  {Smilga}}]{BMS95}%
  \BibitemOpen
  \bibfield  {author} {\bibinfo {author} {\bibfnamefont {D.}~\bibnamefont
  {B{\"o}deker}}, \bibinfo {author} {\bibfnamefont {L.~D.}\ \bibnamefont
  {McLerran}},\ and\ \bibinfo {author} {\bibfnamefont {A.~V.}\ \bibnamefont
  {Smilga}},\ }\bibfield  {title} {\bibinfo {title} {Really computing
  nonperturbative real time correlation functions},\ }\href
  {https://doi.org/10.1103/PhysRevD.52.4675} {\bibfield  {journal} {\bibinfo
  {journal} {Phys. Rev. D}\ }\textbf {\bibinfo {volume} {52}},\ \bibinfo
  {pages} {4675} (\bibinfo {year} {1995})},\ \Eprint
  {https://arxiv.org/abs/hep-th/9504123} {arXiv:hep-th/9504123} \BibitemShut
  {NoStop}%
\bibitem [{\citenamefont {Aarts}\ and\ \citenamefont {Berges}(2002)}]{AB2002}%
  \BibitemOpen
  \bibfield  {author} {\bibinfo {author} {\bibfnamefont {G.}~\bibnamefont
  {Aarts}}\ and\ \bibinfo {author} {\bibfnamefont {J.}~\bibnamefont {Berges}},\
  }\bibfield  {title} {\bibinfo {title} {Classical aspects of quantum fields
  far from equilibrium},\ }\href
  {https://doi.org/10.1103/PhysRevLett.88.041603} {\bibfield  {journal}
  {\bibinfo  {journal} {Phys. Rev. Lett.}\ }\textbf {\bibinfo {volume} {88}},\
  \bibinfo {pages} {041603} (\bibinfo {year} {2002})},\ \Eprint
  {https://arxiv.org/abs/hep-ph/0107129} {arXiv:hep-ph/0107129} \BibitemShut
  {NoStop}%
\bibitem [{\citenamefont {Berges}\ \emph {et~al.}(2004)\citenamefont {Berges},
  \citenamefont {Bors{\'a}nyi},\ and\ \citenamefont {Wetterich}}]{BBW04}%
  \BibitemOpen
  \bibfield  {author} {\bibinfo {author} {\bibfnamefont {J.}~\bibnamefont
  {Berges}}, \bibinfo {author} {\bibfnamefont {S.}~\bibnamefont
  {Bors{\'a}nyi}},\ and\ \bibinfo {author} {\bibfnamefont {C.}~\bibnamefont
  {Wetterich}},\ }\bibfield  {title} {\bibinfo {title} {Prethermalization},\
  }\href {https://doi.org/10.1103/PhysRevLett.93.142002} {\bibfield  {journal}
  {\bibinfo  {journal} {Phys. Rev. Lett.}\ }\textbf {\bibinfo {volume} {93}},\
  \bibinfo {pages} {142002} (\bibinfo {year} {2004})},\ \Eprint
  {https://arxiv.org/abs/hep-ph/0403234} {arXiv:hep-ph/0403234} \BibitemShut
  {NoStop}%
\bibitem [{\citenamefont {Micha}\ and\ \citenamefont {Tkachev}(2004)}]{MT04}%
  \BibitemOpen
  \bibfield  {author} {\bibinfo {author} {\bibfnamefont {R.}~\bibnamefont
  {Micha}}\ and\ \bibinfo {author} {\bibfnamefont {I.~I.}\ \bibnamefont
  {Tkachev}},\ }\bibfield  {title} {\bibinfo {title} {Turbulent
  thermalization},\ }\href {https://doi.org/10.1103/PhysRevD.70.043538}
  {\bibfield  {journal} {\bibinfo  {journal} {Phys. Rev. D}\ }\textbf {\bibinfo
  {volume} {70}},\ \bibinfo {pages} {043538} (\bibinfo {year} {2004})},\
  \Eprint {https://arxiv.org/abs/hep-ph/0403101} {arXiv:hep-ph/0403101}
  \BibitemShut {NoStop}%
\bibitem [{\citenamefont {Wetterich}(2017)}]{CWFI}%
  \BibitemOpen
  \bibfield  {author} {\bibinfo {author} {\bibfnamefont {C.}~\bibnamefont
  {Wetterich}},\ }\bibfield  {title} {\bibinfo {title} {Fermions as generalized
  {I}sing models},\ }\href {https://doi.org/10.1016/j.nuclphysb.2017.02.012}
  {\bibfield  {journal} {\bibinfo  {journal} {Nucl. Phys. B}\ }\textbf
  {\bibinfo {volume} {917}},\ \bibinfo {pages} {241} (\bibinfo {year}
  {2017})},\ \Eprint {https://arxiv.org/abs/1612.06695} {arXiv:1612.06695
  [cond-mat.stat-mech]} \BibitemShut {NoStop}%
\bibitem [{\citenamefont {Wetterich}(2026{\natexlab{d}})}]{CWCPT}%
  \BibitemOpen
  \bibfield  {author} {\bibinfo {author} {\bibfnamefont {C.}~\bibnamefont
  {Wetterich}},\ }\bibfield  {title} {\bibinfo {title} {Complex wave functions,
  {CPT} and quantum field theory for classical generalized {I}sing models},\
  }\href {https://doi.org/10.1016/j.nuclphysb.2025.117204} {\bibfield
  {journal} {\bibinfo  {journal} {Nucl. Phys. B}\ }\textbf {\bibinfo {volume}
  {1022}},\ \bibinfo {pages} {117204} (\bibinfo {year} {2026}{\natexlab{d}})},\
  \Eprint {https://arxiv.org/abs/2505.24392} {arXiv:2505.24392 [quant-ph]}
  \BibitemShut {NoStop}%
\end{thebibliography}%

\end{document}